\documentclass[a4paper,aps,nofootinbib,11pt,titlepage,nofootinbib]{revtex4-1}
\usepackage{amsfonts,color}
\usepackage{graphicx,calc,epsfig}
\usepackage{multirow}
\usepackage{color}

\newcounter{mnotecount}[section]

\def\tyng(#1){\hbox{\tiny$\yng(#1)$}}	
\def\smng(#1){\hbox{\small$\yng(#1)$}}	
\newcommand{\be}{\nopagebreak[3]\begin{equation}}
\newcommand{\ee}{\end{equation}}
\newcommand{\ba}{\nopagebreak[3]\begin{eqnarray}}
\newcommand{\ea}{\end{eqnarray}}

\def\Vrule{\vrule width 1.2pt}

\def\doublyso{\kern-.5em\smash{\vrule height \rowcount em depth .2em}\hidewidth}

\usepackage{hyperref}
\usepackage{cleveref}[2012/02/15]
\usepackage{psfrag}
\usepackage{vmargin}
\usepackage{scrextend}
\usepackage{epsfig}
\usepackage{amsfonts}
\usepackage{young}
\usepackage{amsmath}
\usepackage{amsmath,amssymb}
\usepackage{blkarray}
\usepackage{mathrsfs}
\usepackage{bbm}
\usepackage{amsfonts}
\usepackage{bm}
\usepackage[vcentermath,enableskew]{youngtab}
\usepackage{amsfonts}
\usepackage{youngtab}
\usepackage{young}
\usepackage[margin=1in]{geometry}
\usepackage{graphics,graphicx,color,calc,epsfig}
\usepackage{euscript}
\newcommand{\beq}{\begin{equation}}
\newcommand{\eeq}{\end{equation}}
\newcommand{\beqa}{\begin{eqnarray}}
\newcommand{\eeqa}{\end{eqnarray}}
\newcommand{\nn}{\nonumber}

\newcommand{\ra}[1]{\mbox{{\raisebox{1.14em}{#1}}}}
\newcommand{\rar}[1]{\mbox{{\raisebox{2.28em}{#1}}}}
\newcommand{\rarr}[1]{\mbox{{\raisebox{3.42em}{#1}}}}
\newcommand{\ray}[1]{\mbox{{\raisebox{-.05em}{#1}}}}

\setpapersize{USletter}
\usepackage{fancyhdr}
\setmarginsrb{25mm}{12mm}{25mm}{12mm}{12pt}{11mm}{2pt}{20mm}
\newcommand*{\TitleFont}{%
      \usefont{\encodingdefault}{\rmdefault}{b}{n}%
      \fontsize{15}{20}%
      \selectfont}
    \usepackage[raggedright]{titlesec}
    \renewcommand \thesection{\arabic{section}.}
       
\renewcommand
 \thesubsection{\arabic{section}.\arabic{subsection}.}
\titleformat{\section}{\Large\bfseries}{\makebox[7pt][l]{\thesection}}{12pt}{}
\titleformat{\subsection}{\large\bfseries}{\makebox[14pt][l]{\thesubsection}}{12pt}{}
\renewcommand{\thesection}{\arabic{section}}
\renewcommand{\thesubsection}{\arabic{section}.\arabic{subsection}.}

\renewcommand{\theequation}{\thesection.\arabic{equation}}
\makeatletter \@addtoreset{equation}{section}

\newcommand{\appendices}{\section*{Appendix}\setcounter{subsection}{0} \setcounter{equation}{0}
\renewcommand{\thesubsection}{\Alph{subsection}.}
\renewcommand{\theequation}{\thesubsection\arabic{equation}}}

\begin{document}
\title{\TitleFont Quantum Hall Effect on the Grassmannians $\mathbf{Gr}_2(\mathbb{C}^N)$}
\author{\large F. Ball\i\footnote{fatih.balli@metu.edu.tr}, A. Behtash\footnote{alireza.behtash@metu.edu.tr}, S. K\"urk\c{c}\"uo\u{g}lu\footnote{kseckin@metu.edu.tr}, G. \"{U}nal\footnote{ugonul@metu.edu.tr} \vadjust{\vskip .5cm \vskip 0pt}}

\affiliation{\large Department of Physics, Middle East Technical University \\
Dumlupinar Boulevard, 06800 \\
Ankara, Turkey
\\}
\vbox{\vskip 5em}
\
\begin{abstract}

\noindent Quantum Hall Effects (QHEs) on the complex Grassmann manifolds $\mathbf{Gr}_2(\mathbb{C}^N)$ are formulated. We set up the Landau problem in $\mathbf{Gr}_2(\mathbb{C}^N)$ and solve it using group theoretical techniques and provide the energy spectrum and the eigenstates in terms of the $SU(N)$ Wigner ${\cal D}$-functions for charged particles on $\mathbf{Gr}_2(\mathbb{C}^N)$ under the influence of abelian and non-abelian background magnetic monopoles or a combination of these thereof. In particular, for the simplest case of $\mathbf{Gr}_2(\mathbb{C}^4)$ we explicitly write down the $U(1)$ background gauge field as well as the single and many-particle eigenstates by introducing the Pl\"{u}cker coordinates and show by calculating the two-point correlation function that the Lowest Landau Level (LLL) at filling factor $\nu =1$ forms an incompressible fluid. Our results are in agreement with the previous results in the literature for QHE on ${\mathbb C}P^N$ and generalize them to all $\mathbf{Gr}_2(\mathbb{C}^N)$ in a suitable manner. Finally, we heuristically identify a relation between the $U(1)$ Hall effect on $\mathbf{Gr}_2(\mathbb{C}^4)$ and the Hall effect on the odd sphere $S^5$, which is yet to be investigated in detail, by appealing to the already known analogous relations between the Hall effects on ${\mathbb C}P^3$ and ${\mathbb C}P^7$ and those on the spheres $S^4$ and $S^8$, respectively.

\end{abstract}

\maketitle

\section{Introduction}

Sometime ago Hu and Zhang introduced a $4$-dimensional generalization of the quantum Hall effect (QHE) \cite{Hu-Zhang}. They have formulated  and solved the Landau problem on $S^4$ for fermions carrying an additional $SU(2)$ degree of freedom and under the influence of an $SU(2)$ background gauge field. For the multi-particle problem in the lowest Landau level (LLL) with filling factor $\nu=1$, it turns out that in the thermodynamic limit, a finite spatial density is achieved only if the particles are in infinitely large irreducible representations of $SU(2)$ (i.e. they carry infinitely large number of $SU(2)$ internal degrees of freedom). In this limit, two-point density correlation function immediately indicates incompressibility property of this $4$-dimensional quantum Hall liquid, as the probability of finding two particles at a distance much shorter than the magnetic length in this model approaches to zero. Appearance of massless chiral bosons at the edge of a $2$-dimensional quantum Hall droplet \cite{Ezawa, Halperin, Wen, Stone} also generalizes to this setting. Nevertheless, it is found that among the edge excitations of this $4$-dimensional quantum Hall droplet not only photons and gravitons but also other massless higher spin states occur. The latter is essentially due to the presence of a large number of $SU(2)$ degrees of freedom attached to each particle and, as such, it is not a desirable feature of the model. In a subsequent article \cite{Bernevig-EFT}, two equivalent effective Chern Simons (CS) field theory descriptions, an abelian CS theory in $6+1$-dimensions and a $SU(2)$ non-abelian CS theory in $4+1$-dimensions, of the quantum Hall droplet on $S^4$ have been constructed as generalizations of the well-known Chern-Simons-Landau-Ginzburg (CSLG) model for  fractional QHE which successfully captures the long wavelength limit structure of quantum Hall fluid as a topological field theory \cite{Zhang-Hansson-Kivelson, Ezawa}. 

Other developments ensued the ground-breaking work of Hu and Zhang. Several authors have addressed other higher-dimensional generalizations of QHE to a variety of manifolds including complex projective spaces ${\mathbb C}P^{N}$, $S^8$, $S^3$, the Flag manifold $\frac{SU(3)}{U(1) \times U(1)}$, as well as quantum Hall systems based on higher dimensional fuzzy spheres \cite{Karabali-Nair-1, Bernevig-S8, Nair-Daemi, Jellal, Hasebe}. Of particular interest to us is the work of Nair and Karabali on the formulation of QHE problem on ${\mathbb C}P^{N}$ \cite{Karabali-Nair-1}. These authors solve the Landau problem on ${\mathbb C}P^{N}$ by appealing to the coset realization of ${\mathbb C}P^{N}$ over $SU(N+1)$ and performing a suitable restriction of the Wigner ${\cal D}$-functions on the latter. In this manner, wave functions for charged particles under the influence of both $U(1)$ abelian and/or non abelian $SU(N)$ gauge backgrounds are obtained as sections of $U(1)$ and/or $SU(N)$ bundles over ${\mathbb C}P^{N}$. This formulation simultaneously permits the authors to give the energy spectrum of the LL, where the degeneracy in each LL is identified with the dimension of the irreducible representation (IRR) to which the aforementioned restricted Wigner ${\cal D}$-functions belong. An important feature of these results is that the spatial density of particles remains finite without the need for infinitely large internal $SU(N)$ degrees of freedom, contrary to the situation encountered for the Hall effect on $S^4$. It also turns out that there is a close connection between the Hall effects on ${\mathbb C}P^3$ and ${\mathbb C}P^7$ with abelian backgrounds and those on the spheres $S^4$ and $S^8$ with $SU(2)$ and $SO(8)$ backgrounds, respectively \cite{Karabali-Nair-1, Bernevig-S8, Hasebe}. Effective actions for the edge dynamics in the limit of large number of fermions at the LLL for $\nu =1$ on ${\mathbb C}P^{N}$ were obtained also by Nair and Karabali in \cite{Nair2, Nair-Karabali-Effective2} for abelian and non-abelian backgrounds respectively, and on $S^3$ with a non-abelian background, which is taken as the spin connection, by Nair and Randjbar-Daemi \cite{Nair-Daemi}. These theories involve either abelian bosonic fields or they are higher dimensional generalizations of gauged WZW models, which are chiral in a sense related to the geometry of these spaces. These investigations reveal that the effective edge action for the QHE on $S^4$ obtained from that of ${\mathbb C}P^{3}$ does not describe a relativistic field theory, although there are states which do satisfy the relativistic dispersion relation. Nevertheless, these models possess several features which make them interesting in their own right and worthy for further investigations.
 
Complex Grassmannians $\mathbf{Gr}_k(\mathbb{C}^N)$ are generalizations of complex projective spaces ${\mathbb C}P^{N}$, which share many of their nice features, such as being a K\"ahler manifold. Several of these features are effectively captured by their so-called the Pl\"{u}cker embedding into ${\mathbb C}P^{\tbinom{N}{k}-1}$. For the case $k=2$, to which we will be restricting ourselves in this paper, the Pl\"{u}cker embedding describes $\mathbf{Gr}_k(\mathbb{C}^N)$ as a projective algebraic hypersurface in ${\mathbb C}P^{N}$. For $\mathbf{Gr}_k(\mathbb{C}^N)$ this is the well-known Klein Quadric in ${\mathbb C}P^{5}$ \cite{Ward-Wells}. The developments summarized above and the intriguing geometry of Grassmannian manifolds motivates us to take up the formulation of the QHE problem on the Grassmannians $\mathbf{Gr}_2(\mathbb{C}^N)$. Using group theoretical techniques, we solve the Landau problem on $\mathbf{Gr}_2(\mathbb{C}^N)$ and provide the energy spectrum and the eigenfunctions in terms of $SU(N)$ Wigner ${\cal D}$-functions for charged particles on $\mathbf{Gr}_2(\mathbb{C}^N)$ under the influence of abelian and/or non-abelian background magnetic monopoles, where the latter are obtained as sections of bundles over $\mathbf{Gr}_2(\mathbb{C}^N)$. In section 3, we first treat the simplest and perhaps the more interesting case of $\mathbf{Gr}_2(\mathbb{C}^4)$, where the solution for the most general case of non-zero $U(1)$ and $SU(2) \times SU(2)$ backgrounds are given. In particular, we show that at the LLL with $\nu =1$, finite spatial densities are obtained at finite $SU(2) \times SU(2)$ internal degrees of freedom in agreement with the results of \cite{Karabali-Nair-1}. In section 4, we generalize these results to all $\mathbf{Gr}_2(\mathbb{C}^N)$. The local structure of the solutions on $\mathbf{Gr}_2(\mathbb{C}^4)$ in the presence of $U(1)$ background gauge field is presented in section 5. There we give the single and multi-particle wave functions by introducing the Pl\"{u}cker coordinates and show by calculating the two-point correlation function that the LLL at filling factor $\nu =1$ forms an incompressible fluid. The $U(1)$ gauge field, its associated field strength and their properties are illustrated using the differential geometry on $\mathbf{Gr}_2(\mathbb{C}^4)$. We also briefly comment on the generalization of this local formulation to all $\mathbf{Gr}_2(\mathbb{C}^N)$.

As we have noted earlier, the QHE problem on ${\mathbb C}P^{3}$ with $U(1)$ is special, because of its close connection to the QHE on $S^4$. Already ${\mathbb C}P^{3}$ has the form $S^4 \times S^2$ locally, but in fact ${\mathbb C}P^{3}$ is the projective twistor space, and it forms a non-trivial fiber bundle over $S^4$ with $S^2$ fibers \cite{Karabali-Nair-1, Ward-Wells}. Thus, another motivation for our work comes from the twistor correspondence between and $\mathbb{C}P^3$ and $\mathbf{Gr}_2(\mathbb{C}^4) \equiv F_2$. The latter is also a twistor manifold and together they  
form the base spaces for the double fibration from the Flag manifold $F_{12}$. We hope that our work may be taken as a first step towards an extensive study to reveal possible twistor correspondences between the QHE on these manifolds and also between QHE formulations on similarly related twistor spaces such as $F_{13}$ and $F_2$.

We discuss our conclusions and some possible future directions of research in section 6. In this section, we also discuss a heuristical correspondence between the $U(1)$ Hall effect on $\mathbf{Gr}_2(\mathbb{C}^4)$ and the Hall effect on the odd sphere $S^5$. An independent treatment of the latter, apart form our proposed suggestion in section 6, is still missing in the literature, while we think that it can be formulated by generalizing the results of \cite{Nair-Daemi} on $S^3$.

\section{Review of QHE on $\mathbb{C}P^1$ and $\mathbb{C}P^2$}

In this section, we provide a short account of the formulation of quantum Hall problem on  $\mathbb{C}P^1$ and $\mathbb{C}P^2$ for the purposes of orienting the developments in the subsequent sections and making the exposition self-contained. The formulation of the QHE on $\mathbb{C}P^1 \equiv S^2$ is originally due to Haldane \cite{Haldane}. Karabali and Nair \cite{Karabali-Nair-1} have provided a reformulation of QHE on  $\mathbb{C}P^1$ in a manner that is adaptable to formulate QHE on  $\mathbb{C}P^N$. Here, we closely follow the discussion of \cite{Karabali-Nair-1} and while at it we provide the Young diagram techniques for handling the QHE problem on $\mathbb{C}P^2$. In section 3 and 4 we employ the latter to transparently handle the branching of the IRR of $SU(N)$ under the relevant subgroups appearing in the coset realizations of $\mathbf{Gr}_2(\mathbb{C}^N)$.

Landau problem on $\mathbb{C}P^1$ can be viewed as electrons on a two-sphere under the influence of a Dirac monopole sitting at the center. Our task is to construct the Hamiltonian for a single electron under the influence of a monopole field. 
To this end, let us first point out that by the Peter-Weyl theorem the functions on the group manifold of $SU(2) \equiv S^3$ may be expanded in terms of the Wigner-${\cal D}$ functions $\mathcal{D}^{(j)}_{L_3R_3}(g)$ where $g$ is an $SU(2)$ group element and $j$ is an integral or a half-odd integral number labeling the IRR of $SU(2)$. The subscripts $L_3$ and $R_3$ are the eigenvalues of the third component of the left- and right-invariant vector fields on $SU(2)$\footnote{Throughout the article we sometimes denote the left and right invariant vector fields of $SU(N)$ and their eigenvalues by $L_i$ and $R_i$, respectively, which one is meant will be clear from the context.}. The left- and right-invariant vector fields on $SU(2)$ satisfy
\be
\lbrack L_i \,, L_j \rbrack = - \varepsilon_{ijk} L_k \,, \quad \lbrack R_i \,, R_j \rbrack =  \varepsilon_{ijk} R_k \,, \quad \lbrack L_i \,, R_j \rbrack = 0 \,.
\ee

The harmonics as well as sections of bundles over $\mathbb{C}P^1$ may be obtained from the Wigner-${\cal D}$ functions on $SU(2)$ by a suitable restriction of the latter. The coset realization of $\mathbb{C}P^1$ is 
\begin{equation}
\mathbb{C}P^1 \equiv S^2=\frac{SU(2)}{U(1)}.
\end{equation} 
This implies that the sections of $U(1)$ bundle over $\mathbb{C}P^1$ should fulfill
\be
{\cal D} (g e^{i R_3 \theta}) = e^{i \frac{n}{2} \theta}  {\cal D} (g) \,,
\label{eq:secofbundle}
\ee 
where $n$ is an integer. This condition is solved by the functions of the form $\mathcal{D}^{(j)}_{L_3 \frac{n}{2}}(g)$. In fact, the eigenvalue $\frac{n}{2}$ of $R_3$ corresponds to the strength of the Dirac monopole at the center of the sphere and $\mathcal{D}^{(j)}_{L_3 \frac{n}{2}}(g)$ are the desired wavefunctions as will be made clear shortly. In particular, $\mathcal{D}^{(j)}_{L_3 0}(g)$ correspond to the spherical harmonics on $S^2$, which are the wavefunctions for electrons on a sphere with zero magnetic monopole background.

In the presence of a magnetic monopole field $B$, the Hamiltonian must involve covariant derivatives whose commutator is proportional to the magnetic field. Let us take this commutator as $\lbrack D_+ \,, D_- \rbrack = B$. It is now observed that the covariant derivatives $D_\pm$ 
may be identified by the right invariant vector fields $R_\pm = R_1 \pm i R_2$, as
\be
D_\pm = \frac{1}{\sqrt{2}\ell} R_\pm \,,
\ee
where $\ell$ denotes the radius of the sphere. Noting that $\lbrack R_+ \,, R_- \rbrack = 2 R_3$, for the eigenvalue $\frac{n}{2}$ of $R_3$ we have
\be
B = \frac{n}{2\ell^2} \,
\ee
for the magnetic monopole with the strength $\frac{n}{2}$ in accordance with the Dirac quantization condition. The associated magnetic flux through the sphere is $2 \pi n$.

The Hamiltonian may be expressed as 
\beqa
H &=& \frac{1}{2M}(D_{+}D_{-} + D_{-}D_{+}) \nn \\
&= & \frac{1}{2M \ell^2} (\sum^3_{i=1} R^2_i - R^2_3) \,,
\eeqa
where $M$ is the mass of the particle. We have that $\sum_{i=1}^3R^2_i=\sum_{i=1}^3L^2_i=j(j+1)$.  In order to guarantee that $\frac{n}{2}$ occurs as one of the possible eigenvalues of $R_3$, we need to have $j=\frac{1}{2}n+q$ where $q$ is an integer. The spectrum of the Hamiltonian reads 
\begin{eqnarray}
E_{q,n}&=&\frac{1}{2M \ell^2} \left ( (\frac{n}{2} + q)(\frac{n}{2}+q+1)-\frac{n^2}{4} \right ) \nn \\
&=&\frac{B}{2M }(2q+1)+\frac{q(q+1)}{2M\ell^2}  
\label{eq:spec1}   
\end{eqnarray}
The associated eigenfunctions are $\mathcal{D}^{(j)}_{L_3 \frac{n}{2}}(g)$ as noted earlier. In \eqref{eq:spec1}, $q$ is readly interpreted as the Landau level (LL) index. The ground state, that is the Lowest Landau Level (LLL), is at $q=0$ and has the energy $\frac{B}{2M}$. The LLL is separated from the higher LL by finite energy gaps. 

The degeneracy of the LL are controlled by the left invariant vector fields $L_i$ since they commute with the covariant derivatives $\lbrack L_i \,, D_j \rbrack = 0$. Each LL is $(2j+1=n+1+2q)$-fold degenerate. In other words, there are this many wavefunctions $\mathcal{D}^{(j)}_{L_3 \frac{n}{2}}(g)$ at a given LL with $L_3$ eigenvalues ranging from $-j$ to $j$. 

Local form of the wavefunctions may be written down by picking a suitable coordinate system.  We omit this here and refer the reader to the original literature \cite{Karabali-Nair-1} where this is done in detail. In particular, it is shown in \cite{Karabali-Nair-1} that the LLL form an incompressible liquid by computing the two-point correlation function for the wave-function density. We will address this crucial property of the LLL for our case in section 5.

Let us now briefly turn our attention to the formulation of Landau problem on $\mathbb{C}P^2$. This and its generalization to $\mathbb{C}P^N$ is given in \cite{Karabali-Nair-1}. The coset realization of $\mathbb{C}P^2$ may be written as 
\begin{equation}
\mathbb{C}P^2 \equiv \frac{SU(3)}{U(2)} \sim \frac{SU(3)}{SU(2)\times U(1)} \,.
\end{equation}
Following a similar line of development as in the previous case, we can obtain the harmonics and local sections of bundles over $\mathbb{C}P^2$ from a suitable restriction of the Wigner-${\cal D}$ functions on $SU(3)$. Let $g \in SU(3)$ and let us denote the left- and the right-invariant vector fields on $SU(3)$ by $L_\alpha$ and $R_\alpha$ ($\alpha: 1\,, \cdots \,,8$); they fulfill the Lie algebra commutation relations for $SU(3)$. We can introduce the Wigner-${\cal D}$ functions on $SU(3)$ as 
\begin{equation}
\mathcal{D}^{(p,q)}_{L,L_3,L_8;R,R_3,R_8}(g) \,,
\end{equation}
where $(p,q)$ label the irreducible representations of $SU(3)$, and the subscripts denote the relevant quantum numbers for the left- and right- rotations. In particular, the left and right generators of the $SU(2)$ subgroup are labeled by $L_i$ and $R_i$ ($i:1\,,2 \,,3$) 
and $L_iL_i = L(L+1)$, $R_iR_i = R(R+1)$.

We note that the tangents along $\mathbb{C}P^2$ may be parametrized by the right invariant fields, $R_\alpha$, ($\alpha: 4,5,6,7$). Consequently, the Hamiltonian on $\mathbb{C}P^2$ may be written down as
\beqa
H &=& \frac{1}{2  M \ell^2} \sum^7_{\alpha = 4} R_\alpha^2 \nn \\
&=&\frac{1}{2 M \ell^2} \left (\mathcal{C}_2(p,q)- R (R+1) - R_8^2 \right ) \,,
\label{eq:Hamcp2}
\eeqa
where $\mathcal{C}_2(p,q)$ is the quadratic Casimir of $SU(3)$.

The coset realization of $\mathbb{C}P^2$ implies that there can be both abelian and non-abelian background gauge fields  corresponding to the gauging of the $U(1)$ and $SU(2)$ subgroups, respectively.  

Let us first obtain the wave functions with the $U(1)$ background gauge field. This means that our desired ${\cal D}^{(p,q)}$ should transform trivially under the $SU(2)$, and carry a $U(1)$ charge under the right actions of these groups. In other words, these wave functions must be singlets under $SU(2)$ with $R =0 \,, R_3 = 0$ and a non-zero $R_8$ eigenvalue. We can utilize the Young tableaux to see the branching of the $SU(3)$ IRR satisfying this requirement. The $SU(3)$ IRR labeled by $(p,q)$ may be assigned to a Young tableau with $p$ columns with one box each and $q$ columns with two boxes on each. The branching $SU(3)\supset SU(2)\times U(1)$, which keeps the $SU(2)$ in the singlet representation, is therefore
\begin{align*}
\overbrace{\ra{\yng(2,2)}\rar{$~\cdots$}}^q\overbrace{\rar{\yng(2)}
\rar{$~\cdots$}}^p\rar{$\longrightarrow~$}
\overbrace{\ra{\yng(2,2)}\rar{$~\cdots$}}^{q}\rar{$\otimes~$}\overbrace{\rar{\yng(2)}\rar{$~\cdots$}}^{p}
\end{align*}
where the diagram on l.h.s. of the arrow represents the generic $(p,q)$ IRR of $SU(3)$ and the first diagram on the r.h.s. of the arrow represent the $SU(2)$ IRR, which is singlet in this case. A general formula exits \cite{Bars} for expressing the $U(1)$ charge of the branching $SU(3)\supset SU(2)\times U(1)$ (see equation \ref{eq:charge} for a more general case):
\be
n = \frac{1}{2}(J_1-2J_2) \,, \quad n \in {\mathbb Z} \nn
\ee
where $J_1$ is the number of boxes in the tableau of $SU(2)$ and $J_2$ is the number of boxes in the rightmost tableau in the branching. Thus for the tableaux given above, we conclude that $n = q - p$. In order to fix the relation between $R_8$ eigenvalues and the integer $n$, we use the fundamental representation $(1,0)$ with the generators $\lambda_a$ fulfilling the normalization condition $\text{Tr}(\lambda_{a}\lambda_b)=\frac{1}{2}\delta_{ab}$, and $\lambda_8 = \frac{1}{2 \sqrt{3}} \text{diag} (1,1,-2)$, so that
\begin{equation}
R_8=-\frac{n}{\sqrt{3}} = - \frac{p - q}{\sqrt{3}}.
\end{equation}
It is useful to note that the flux of the $U(1)$ field strength corresponding to the background gauge field is proportional to the number 
$n$. We omit the details of this here and refer the reader to \cite{Karabali-Nair-1}.

The spectrum of the Hamiltonian (\ref{eq:Hamcp2}) may be given as
\begin{equation}
E_{q,n}=\frac{1}{2 M \ell^2} \left ( q (q+n+2)+ n \right )\,,
\label{eq:ecp2}
\end{equation}
where we have used the eigenvalue of the quadratic Casimir $\mathcal{C}_2(p,q)$ of the IRR $(p,q)$, which is 
\begin{equation}
\mathcal{C}_2(p,q) = \frac{1}{3}\left ( p(p+3)+q(q+3)+pq \right ) \,.
\end{equation}
and expressed the energy levels in terms of $q$ and $n$ only. In (\ref{eq:ecp2}), $q$ appears as the Landau level index; the ground state energy may be obtained by setting $q=0$ and that gives LLL energy $E_{LLL}=\frac{n}{2M\ell^2}$. 

The wave functions corresponding to this energy spectrum can be written in terms of the Wigner-${\cal D}$ functions as 
\begin{equation}
\mathcal{D}^{(p,q)}_{L,L_3,L_8;0,0,-\frac{n}{\sqrt{3}}}(g) \,.
\end{equation}

The degeneracy of each Landau level $q$ is given by the dimension of the IRR $(p,q)$, which is
\be
\mbox{dim}(p,q) = \frac{(p+1)(q+1)(p+q+2)}{2} \,. 
\ee
This means that the set of quantum numbers $L,L_3$ and $L_8$ can take $\mbox{dim}(p,q)$ different values. 

It is also useful to note that the case $n=0$ simply reduces the Wigner-${\cal D}$ functions to the harmonics on ${\mathbb C}P^2$, corresponding to the wave functions of a particle on ${\mathbb C}P^2$ with vanishing monopole background.

Consider the case of filling factor $\nu =1$, i.e. each of the LL states is occupied by one fermion. We therefore have that $p=n$, $q=0$ and the number of fermions ${\mathcal N}$ is equal to $\mbox{dim}(n,0) = (n+1)(n+2)/2$. The density of particles $\rho$ is given by
\begin{equation}
\rho=\frac{\mathcal N}{{\text {vol}}({\mathbb C}P^2)} \,,
\end{equation}
where ${\text {vol}}({\mathbb C}P^2)= 8 \pi^2 \ell^4$. In the thermodynamic limit $ \ell \rightarrow \infty$ and ${\mathcal N} \rightarrow \infty$, this yields the finite result
\begin{equation}
\rho=\frac{{\cal N}}{8 \pi^2 \ell^4} \underset{\ell \rightarrow \infty \,, {\mathcal N} \rightarrow \infty}{\longrightarrow} \frac{n^2}{16\pi^2\ell^4} = \left(\frac{B}{2\pi}\right)^2,
\end{equation}
as first discussed in \cite{Karabali-Nair-1}.

The wave functions can be expressed in suitable local coordinates and, taking advantage of these functions, the multi-particle wave-function for the filling factor $\nu =1$ state can immediately be constructed. A straightforward calculation for the two-point correlation function for the wave-function density may be given which signals the incompressibility of the LLL. We refer the reader to \cite{Karabali-Nair-1} for details.

The case of $SU(2)$ and $U(1)$ background gauge fields may be handled as follows. In this case we allow for all possible right $SU(2)$ IRR labeled by spin $R$. It is possible to label $SU(3)$ representations in the form $(p+k,q+k^\prime)$. The branching $SU(3)\supset SU(2)\times U(1)$ may be represented by the Young tableaux 
\begin{align*}
\overbrace{\ra{\yng(2,2)}\rar{$~\cdots$}}^{q+k'}\overbrace{\rar{\yng(2)}
\rar{$~\cdots$}}^{p+k}&\rar{$\longrightarrow~$}
\overbrace{\ra{\yng(2,2)}\rar{$~\cdots$}}^q\overbrace{{\rar{\yng(2)}\rar{$~\cdots$}}}^{k+k'}\rar{$\otimes~$}\overbrace{\rar{\yng(2)}\rar{$~\cdots$}}^{p}\overbrace{{\rar{\yng(2)}\rar{$~\cdots$}}}^{k'}
\\
&\rar{$\longrightarrow~$}
\overbrace{\ra{\yng(2,2)}\rar{$~\cdots$}}^{q+k'-x}\overbrace{{\rar{\yng(2)}\rar{$~\cdots$}}}^{k-k'+2x}\rar{$\otimes~$}\overbrace{\rar{\yng(2)}\rar{$~\cdots$}}^{p}\overbrace{{\rar{\yng(2)}\rar{$~\cdots$}}}^{k'}
\\
&\rar{$\longrightarrow~$}
\overbrace{\ra{\yng(2,2)}\rar{$~\cdots$}}^{q+k'}\overbrace{{\rar{\yng(2)}\rar{$~\cdots$}}}^{k'-k}\rar{$\otimes~$}\overbrace{\rar{\yng(2)}\rar{$~\cdots$}}^{p}\overbrace{{\rar{\yng(2)}\rar{$~\cdots$}}}^{k'}
 \end{align*}
These tableaux represent the maximum, generic and minimum spin $R$-value configurations that can result from the branching, and we have assumed without loss of generality that $k'>k$ and $k\geq x \geq 0$. Here $x$ is an integer introduced to conveniently represent the generic case. From the tableaux, the range of the spin $R$ and $R_8$ eigenvalues may be easily obtained as follows:
\begin{eqnarray}
\label{isospin_range} R&=&\frac{|k-k'|}{2},\cdots,\frac{k+k'}{2}  \\
R_8 &=&\frac{1}{2\sqrt{3}}\left (-2(p-q)+(k-k') \right ) = - \frac{n}{\sqrt{3}}
\end{eqnarray} 
Noting that $n$ is an integer restricts the spin $R$ to integer values.
Spectrum of the Hamiltonian (\ref{eq:Hamcp2}) is now
\begin{eqnarray}
\label{eigenvalueCP2} E &=& \frac{1}{2 M {\ell}^2}( \mathcal{C}_2(p+k,q+k')-R(R+1)-R_8^2) \\
&=&\frac{1}{2 M {\ell}^2} \left ( q^2+q(2k-m+n+2)+n(k+1)+k^2+2k+m^2-m(k+1)-R(R+1) \right ) \nn
\end{eqnarray}
where $k'=k-2m$ and $m$ is an integer. As indicated in (\ref{isospin_range}), there is an interval for the values of $R$. The LLL is obtained when we choose the maximum value for $R$,
\begin{equation}
\label{RmaxCP2} R_{max}=\frac{k+k'}{2}=k-m\,,
\end{equation}
where $m$ should take only integer values within the interval $m=0,\cdots, \frac{k}{2}$ if $k$ is even, and $m = 0,\cdots, \frac{k-1}{2}$ if $k$ is odd. Using (\ref{RmaxCP2}) in (\ref{eigenvalueCP2}), the energy spectrum is expressed as
\begin{equation}
E=\frac{1}{2 M \ell^2} \left(q^2+q(2R+n+m+2)+n(R+m+1)+(R+m)(m+1) \right) \,.
\end{equation} 
For fixed $n$,$R$ we observe from this expression that the LL are controlled by the two integers $q$ and $m$. The LLL is obtained for $q=0$ and $m=0$. 

As discussed in \cite{Karabali-Nair-1}, for pure $SU(2)$ background, to ensure the finiteness of energy eigenvalues $R$ should scale like $R\sim \ell^2$ in the thermodynamic limit. For $\nu =1$ we have ${\cal N} = \dim(R,R) =\frac{1}{2}(R+1)(R+1)(2R+2) $ and this results in a finite density of particles
\begin{equation}
\rho \sim \frac{{\cal N}}{(2R+1)\ell^4} \underset{ \ell \rightarrow \infty \,, {\mathcal N} \rightarrow \infty}{\longrightarrow} \frac{R^3}{2R\ell^4} \,\,.
\end{equation} 
As for the case of both $U(1)$ and $SU(2)$ backgrounds, it is possible to pick either $n$ or $R$  to scale like $ \ell^2$. Taking $n\sim \ell^2$ and $R$ to be finite as $\ell \rightarrow \infty$, gives again a finite spatial density
\begin{equation}
\rho \sim \frac{\dim{(R+n,R)}}{(2R+1)\ell^4} \underset{\ell \rightarrow \infty \,, {\mathcal N} \rightarrow \infty}{\longrightarrow}  \frac{n^2}{4\ell^4} \,,
\end{equation}
for $\nu =1$ with $\dim(R+n,R)=\frac{1}{2}(n+R+1)(R+1)(n+2R+1)$.

\section{Landau Problem on the Grassmannian $\mathbf{Gr}_2(\mathbb{C}^4)$}

\noindent Starting in this section we will consider the quantum Hall problem on the complex Grassmannians $\mathbf{Gr}_2(\mathbb{C}^N)$. In order to set up the Landau problem on $\mathbf{Gr}_2(\mathbb{C}^N)$, it is necessary to list a few facts about the Grassmannians and their geometry.

The complex Grassmannians $\mathbf{Gr}_k(\mathbb{C}^N)$ are the set of all $k$-dimensional linear subspaces of the vector space $\mathbb{C}^N$ with the complex dimension $k(N-k)$. They are smooth and compact complex manifolds and admit K\"ahler structures. Grassmannians are homogeneous spaces and can therefore be realized as the cosets of $SU(N)$ as 
\begin{equation} \mathbf{Gr}_k(\mathbb{C}^N)= \frac{SU(N)}{S[U(N-k)\times U(k)]} \sim \frac{SU(N)}{SU(N-k)\times SU(k)\times U(1)}.\label{eq:1}
\end{equation} 
It is clear from this realization that $\mathbf{Gr}_1(\mathbb{C}^N)\equiv \mathbb{C}P^N$. $\mathbf{Gr}_2(\mathbb{C}^4)$ is therefore the simplest Grassmannian that is not a projective space. The coset space realization of the Grassmannians is the most suitable setting for group theoretical techniques that we will employ to formulate and solve the Landau problem on $\mathbf{Gr}_2(\mathbb{C}^4)$ first and subsequently on all $\mathbf{Gr}_2(\mathbb{C}^N)$.

In order to set up and solve the Landau problem on $\mathbf{Gr}_2(\mathbb{C}^4)$, we contemplate, following the ideas reviewed in the previous section, that $SU(4)$ Wigner $\mathcal{D}$-functions may be suitably restricted to obtain the harmonics and local sections of bundles over $\mathbf{Gr}_2(\mathbb{C}^4)$. Let $g \in SU(4)$ and let us denote the left- and the right-invariant vector fields on $SU(4)$ by $L_\alpha$ and $R_\alpha$ ($\alpha: 1\,, \cdots \,,15$); they fulfill the Lie algebra commutation relations for $SU(4)$. We can introduce the Wigner-${\cal D}$ functions on $SU(4)$ as 
\begin{equation}
g \rightarrow \mathcal{D}^{(p,q,r)}_{L^{(1)}L^{(1)}_{3}L^{(2)}L^{(2)}_{3} L_{15};R^{(1)}R^{(1)}_{3}R^{(2)}R^{(2)}_{3} R_{15}}(g)
\label{eq:4}
\end{equation}
where $(p,q,r)$ are three integers labeling the irreducible representations of $SU(4)$, and the subscripts denote the relevant quantum numbers for the left- and right- rotations. In particular, the left and right generators of $SU(2) \times SU(2)$ subgroup are labeled by 
$L_\alpha \equiv (L^{(1)}_i, L^{(2)}_i)$ and $R_\alpha \equiv (R^{(1)}_i, R^{(2)}_i)$ ($i:1\,,2 \,,3$, $\alpha: 1\,, \cdots \,, 6$) with corresponding $SU(2) \times SU(2)$ quadratic Casimirs ${\mathcal C}_2^L= L^{(1)}(L^{(1)}+1)+ L^{(2)}(L^{(2)}+1)$, ${\mathcal C}_2^R= R^{(1)}(R^{(1)}+1)+ R^{(2)}(R^{(2)}+1)$.

The real dimension of $\mathbf{Gr}_2(\mathbb{C}^4)$ is $8$ and tangents along $\mathbf{Gr}_2(\mathbb{C}^4)$ may be parametrized by the $8$ right invariant fields $R_\alpha$ ($\alpha: 7\,, \cdots\,,14$). Consequently, the Hamiltonian on $\mathbf{Gr}_2(\mathbb{C}^4)$ may be written down as
\beqa
H &=& \frac{1}{2  M \ell^2} \sum^{14}_{\alpha = 7} R_\alpha^2 \nn \\
&=&\frac{1}{2 M \ell^2} \left (\mathcal{C}_2(p,q,r) - {\mathcal C}_2^R - R_{15}^2 \right ) \,,
\label{eq:Hamgr42}
\eeqa
where $\mathcal{C}_2(p,q,r)$ is the quadratic Casimir of $SU(4)$ in the IRR $(p,q,r)$ with the eigenvalue
\begin{equation}
\mathcal{C}_2(p,q,r)=\frac{3}{8}(r^2+p^2)+\frac{1}{2}q^2+\frac{1}{8}(2pr+4pq+4qr+12p+16q+12r).
\label{eq:C2}
\end{equation}
The dimension of the IRR $(p\,, q \,, r)$ is
\be
\mbox{dim}(p\,, q \,, r) = \frac{1}{12} (p+q+2)(p+q+r+3)(q+r+2)(p+1)(q+1)(r+1). 
\label{eq:deg1}
\ee

The coset realization of $\mathbf{Gr}_2(\mathbb{C}^4)$ implies that, there can be both abelian and non-abelian background gauge fields corresponding to the gauging of the $U(1)$ and one or both of the $SU(2)$ subgroups.  We list these as three distinct cases:
\begin{itemize}
\item[{\it i.}] $U(1)$ background gauge fields only
\item[{\it ii.}] $U(1)$ background gauge field and a single $SU(2)$ background gauge field,
\item[{\it iii.}] $U(1)$ background gauge field and $SU(2) \times SU(2)$ background gauge field.
\end{itemize}
It is useful to remark that the second case may be viewed as a certain restriction of the third. We will discuss these matters in detail in what follows. 

Following \cite{Dolan, Balantekin}, it is useful to list a few facts regarding the branching
\begin{equation}
SU(N_1 + N_2) \supset SU(N_1) \times SU(N_2) \times U(1) \,.\label{eq:bsu_fund}
\end{equation}
We can embed $SU(N_1) \times SU(N_2) \times U(1)$ into $SU(N_1 + N_2)$ as 
\begin{equation} 
\left(\begin{array}{cc} 
e^{i N_2 \phi}U_{1} & 0\\
0 & e^{-i N_1 \phi}U_{2}
\end{array}\right) \,,
\label{eq:2}
\end{equation} 
where $U_1 \in SU(N_1)$ and $U_2 \in SU(N_2)$. Let us denote the IRR of $SU(N_1)$ and $SU(N_2)$ with ${\cal J}_1$ and ${\cal J}_2$. We also let $J_a$ be the total number of boxes in the Young tableaux of $SU(N_a)$ $(a:1,2)$. The $U(1)$ charge may thus be expressed 
as 
\beq  
n = \frac{1}{N_1 N_2}(N_2 J_1- N_1 J_2) \,.
\label{eq:charge}
\eeq
Clearly, the IRR of $U(1)$ is fixed by those of the $SU(N_a)$ factors and the IRR content of the subgroup $SU(N_1) \times SU(N_2) \times U(1)$ may be denoted as $({\cal J}_1,{\cal J}_2)_n$. The decomposition of a given IRR ${\cal J}$ of $SU(N_1+N_2)$ under this subgroup is expressed as 
\begin{equation} 
{\cal J} = \bigoplus_{{\cal J}_1,{\cal J}_2}m^{{\cal J}}_{{\cal J}_1,{\cal J}_2}({\cal J}_1,{\cal J}_2)_n \,,
\label{eq:3}
\end{equation} 
where $m^{{\cal J}}_{{\cal J}_1,{\cal J}_2}$ are the multiplicities of the IRR $({\cal J}_1,{\cal J}_2)_n$ occurring in the direct sum. Further details may be found in the references \cite{Dolan, Balantekin} and in the original article of Hagen and Macfarlane \cite{Hagan}.

\subsection{$U(1)$ gauge field background}

\noindent For the QHE problem on $\mathbf{Gr}_2(\mathbb{C}^4)$ we are concerned with the branching
\begin{equation}
SU(4) \supset SU(2) \times SU(2) \times U(1) \,.\label{eq:bsu4}
\end{equation}
Obtaining the wave functions with the $U(1)$ background gauge field, requires us to restrict ${\cal D}^{(p,q,r)}$ in such a way that they transform trivially under the right action of $SU(2) \times SU(2)$, and carry a right $U(1)$ charge, that is, they should be singlets under $SU(2) \times SU(2)$ with ${\mathcal C}_2^R = 0$ eigenvalue and a non-zero $R_{15}$ eigenvalue. 

We can utilize the Young tableaux to see the branching of the $SU(4)$ IRR fulfilling this requirement. The $SU(4)$ IRR labeled by $(p,q,r)$ may be denoted as a Young tableau with $p$ columns with one box on each, $q$ columns with two boxes on each, and $r$ columns with three boxes on each. The branching \eqref{eq:bsu4}, which keeps the $SU(2) \times SU(2)$ in the singlet representation, is therefore
\begin{align}
\overbrace{\ray{\yng(2,2,2)}\rar{$~\cdots$}}^r\overbrace{\ra{\yng(2,2)}\rar{$~\cdots$}}^q\overbrace{\rar{\yng(2)}
\rar{$~\cdots$}}^p\rar{$\longrightarrow~$} \overbrace{\ra{\yng(2,2)}\rar{$~\cdots$}}^{r+q_1}\rar{$\otimes~$}\overbrace{\ra{\yng(2,2)}\rar{$~\cdots$}}^{q_2}\underbrace{\overbrace{{\ra{\yng(2,2)}\rar{$~\cdots$}}}^r}_p
\label{yg1}
\end{align}
where we have introduced the splitting $q=q_1 + q_2$ in the representation in order to handle the partition of columns labeled by $q$ in the branching. It is important to realize that in the last row of the $SU(4)$ representation there are $r$ (fully symmetrized) boxes, which are moved as a whole under this branching to the second slot in the r.h.s. and the trivial representation of $SU(2)\times SU(2)$ is obtained if and only if $p$ is equal to $r$. Otherwise, we have a nontrivial representation for the second $SU(2)$ in the branching (\ref{eq:bsu4}).

Using the formula (\ref{eq:charge}) we compute the $U(1)$ charge as
\begin{equation}
n = \frac{1}{2} \left ( (2r+2q_1)-(p+r+2q_2) \right ) = q_1 - q_2 \,,
\label{eq:7}
\end{equation}
where we have used $p=r$. 

In order to fix the relation between the eigenvalues of $R_{15}$ and the $U(1)$ charge $n$, we need to use the $6$-dimensional fundamental representation $(0,1,0)$ (Young tableaux: $\tiny{\yng(1,1)}$) of $SU(4)$. As opposed to $\mathbb{C}P^3 \approx SU(4)/SU(3) \times U(1)$ where the branching of the $4$-dimensional  representations (i.e. $(1,0,0)$ and $(0,0,1)$) of $SU(4)$ contain singlets of $SU(3)$, in the present case however, the smallest $SU(4)$ IRR containing the singlet of $SU(2) \times SU(2)$ is $(0,1,0)$ and it has the branching
\begin{equation}
\smng(1,1)~\ \longrightarrow \ ~
\bigg(\cdot~\otimes~\smng(1,1)~\bigg)_{-1}\oplus~\bigg(~\smng(1,1)~\otimes~\cdot\bigg)_1
~\oplus~\bigg(~\smng(1)~\otimes~\smng(1)~\bigg)_0,
\end{equation}
where subscripts show the charge \eqref{eq:7}. Taking the generators $\lambda_a$ of $SU(4)$ fulfilling the normalization condition $\text{Tr}(\lambda_{a}\lambda_b)=\frac{1}{2}\delta_{ab}$, in one of the 4-dimensional IRR $( (1,0,0) \, \, \mbox{or} \, \, (0,0,1))$, it is possible to show that in the $6$-dimensional IRR\footnote{Generalizing this result to the $\frac{N(N-1)}{2}$-dimensional representations of $SU(N)$ is used in the subsequent sections. A proof is provided in appendix A.} $(0,1,0)$
\be
R_{15} = \frac{1}{\sqrt{2}} \text{diag}(0,0,0,0,-1,1), 
\label{eq:chargeU(1)}
\ee
and therefore we in general have 
\be
R_{15} = \frac{n}{\sqrt{2}} = \frac{q_1 - q_2}{\sqrt{2}}.
\label{eq:r15}
\ee

It is now easy to give the energy spectrum corresponding to the Hamiltonian (\ref{eq:Hamgr42}), using (\ref{eq:C2}), $p=r$, $R_{15}$ taking the value in (\ref{eq:r15}) and ${\mathcal C}_2^R=0$:
\be
E = \frac{1}{2 M \ell^2} \left ( p^2+3p+np+2q_2^2+4q_2 + 2 p q_2+2n(1+q_2) \right) \,.
\ee
The LLL energy at a fixed monopole background $n$ is obtained for $q_2 = p = 0$ and it is
\be
E_{LLL} = \frac{n}{ M \ell^2} = \frac{2 B}{ M} \,,
\label{eq:LLL-energy}
\ee
with the degeneracy $\dim (0\,, n \,, 0) = \frac{1}{12}(n+1)(n+2)^2(n+3)$. In (\ref{eq:LLL-energy}), $B = \frac{n}{2 \ell^2}$ is the field strength of the $U(1)$ magnetic monopole. The gauge field associated to $B$ and related matters will be discussed in section 5.

The wave functions corresponding to this energy spectrum can be written in terms of the Wigner-${\cal D}$ functions as 
\be
\mathcal{D}^{(p\,, q_1 + q_2\,, p)}_{L^{(1)}L^{(1)}_{3}L^{(2)}L^{(2)}_{3} L_{15}\, ; \, 0\,,0\,, 0\,, 0\,, \frac{n}{\sqrt{2}}}(g)
\equiv \mathcal{D}^{(p\,, \lbrack \frac{q+n}{2} \rbrack + \lbrack \frac{q-n}{2} \rbrack \,, p)}_{L^{(1)}L^{(1)}_{3}L^{(2)}L^{(2)}_{3} L_{15}\, ; \, 0\,,0\,, 0\,, 0\,, \frac{n}{\sqrt{2}}}(g) \,.
\ee
The degeneracy of each Landau level is given by the dimension of the IRR $(p\,, q \,, p)$ in equation \eqref{eq:deg1}. This means that the set of left quantum numbers $\lbrace L^{(1)}\,, L^{(1)}_{3}\,, L^{(2)}\,, L^{(2)}_{3} \,,L_{15} \rbrace$ can take on $\mbox{dim}(p\,, q_1 + q_2 \,, p)$ different values as a set. 

For the many-body fermion problem in which all the states of LLL are filled with the filling factor $\nu=1$, in the thermodynamic limit $\ell \rightarrow \infty, {\cal N} \rightarrow \infty$ we obtain a finite spatial density of particles
\begin{equation}
\rho=\frac{\cal N}{\frac{\pi^4 \ell^8}{12}} \underset{\ell \rightarrow \infty \,, {\cal N} \rightarrow \infty}{\longrightarrow}  \frac{n^4}{\pi^4 \ell^8}=\left(\frac{2B}{\pi}\right)^4 \,,
\end{equation}
where we have used ${\cal N} = \mbox{dim}(0\,, n \,,0) = \frac{1}{12} (n+1)(n+2)^2(n+3)$ for the number of fermions in the LLL with $\nu=1$, and\footnote{It may be useful to state that this volume is computed with the help of the repeated iteration of (special) unitary group manifolds in terms of the odd dimensional spheres,
\beqa
SU(N)&\approx& \frac{SU(N)}{SU(N-1)}\times \frac{SU(N-1)}{SU(N-2)}\times\cdots \times \frac{SU(3)}{SU(2) } \times SU(2) \nonumber \\
&\cong& S^{2N-1}\times S^{2N-3}\times \cdots \times S^{5} \times S^{3},
\eeqa
(for $N\geq3$) where $\approx$ means ``locally equal to'' and $\cong$ indicates isomorphism. Considering this local expression we can expand all the special unitary groups in \eqref{eq:1} and employ the volume formula for spheres to obtain an approximation for the volume of the Grassmannians \cite{Fuji}, namely,
\be
{\text {vol}}(\mathbf{Gr}_k(\mathbb{C}^N)) = \frac{1!2!\cdots (k-1)!}{(N-1)!(N-2)!\dots (N-k)!} (\pi\ell^2)^{k(N-k)},\label{volume}
\ee
which produces the factor $\tfrac{1}{12}$ for $k=2$ and $N=4$. This factor is in general subject to change upon using other methods. Since this is immaterial for our purposes, we will stick to the approximation \eqref{volume} throughout this paper.} ${\text {vol}}(\mathbf{Gr}_2(\mathbb{C}^4))=\frac{\pi^4 \ell^8}{12}$. 
 
We note that the case $n=0$ simply reduces the Wigner-${\cal D}$ functions to the harmonics on $\mathbf{Gr}_2(\mathbb{C}^4)$ corresponding to the wave functions of a particle on $\mathbf{Gr}_2(\mathbb{C}^4)$ with vanishing monopole background.

It is possible to interchange the Young tableaux of the two $SU(2)$'s in (\ref{yg1}). This flips the sign of the $U(1)$ charge, $n \rightarrow -n$; in the formulas for the energy and degeneracy, etc, this fact can be compensated by substituting $|n|$ for $n$.

In section 5 we give the single and many-particle wave functions  (for the filling factor $\nu =1$ state) in terms of the Pl\"ucker coordinates for $\mathbf{Gr}_2(\mathbb{C}^4)$ and use the latter to obtain the two-point correlation function for the wave-function density signaling the incompressibility of the LLL.  An account of the $U(1)$ gauge field is also provided for illustrative purposes.

\subsection{Single $SU(2)$ gauge field and $U(1)$ gauge field background}

In this case, we need to restrict to ${\cal D}^{(p,q,r)}$, which transform as a singlet under one or the other $SU(2)$ in the right action of $SU(2) \times SU(2)$, and carry a $U(1)$ charge. Therefore, we have a range of possibilities within the branching \eqref{eq:bsu4} as given in the following Young tableaux decomposition:
\begin{align}
 \overbrace{\ray{\yng(2,2,2)}\rar{$~\cdots$}}^r\overbrace{\ra{\yng(2,2)}\rar{$~\cdots$}}^q\overbrace{\rar{\yng(2)}
\rar{$~\cdots$}}^p&\rar{$\longrightarrow~$} \overbrace{\ra{\yng(2,2)}\rar{$~\cdots$}}^{r+q_1}\rar{$\otimes~$}\overbrace{\ra{\yng(2,2)}\rar{$~\cdots$}}^{q_2}\overbrace{{\rar{\yng(2)}\rar{$~\cdots$}}}^{p+r}
\\&\rar{$\longrightarrow~$} \overbrace{\ra{\yng(2,2)}\rar{$~\cdots$}}^{r+q_1}\rar{$\otimes~$}\overbrace{\ra{\yng(2,2)}\rar{$~\cdots$}}^{q_2}\overbrace{{\ra{\yng(2,2)}\rar{$~\cdots$}}}^{x}\overbrace{\rar{\yng(2)}\rar{$~\cdots$}}^{p+r-2x}
\\&\rar{$\longrightarrow~$} \overbrace{\ra{\yng(2,2)}\rar{$~\cdots$}}^{r+q_1}\rar{$\otimes~$}\overbrace{\ra{\yng(2,2)}\rar{$~\cdots$}}^{q_2}\underbrace{{\ra{\yng(2,2)}}\rar{$~\cdots$}}_r\overbrace{\rar{\yng(2)}\rar{$~\cdots$}}^{p-r} \label{yng1}
 \end{align}
We have assumed that $p > r$ and split $q_1+q_2 = q$. We have introduced the integer $x$ ($0 \leq x \leq r$) to conveniently represent the generic case. From the tableaux, $R_{15}$ eigenvalues may be easily obtained as:
\be
n = \frac{1}{2} \left ( 2 ( q_1 - q_2 ) - (p-r) \right ) \,,
\label{eq:chgn}
\ee
and we observe that the first $SU(2)$ in the branching remains a singlet while the second may take on values over a range;
\be
R^{(1)} = 0 \,, \quad R^{(2)}= \frac{p-r}{2} \,, \cdots \,, \frac{r + p}{2} \,.
\label{eq:spins1}
\ee
Since $n$ is an integer, we must have that $p-r$ is an even integer. This condition restricts the spin $R^{(2)}$ to integer values. 

Using ${\mathcal C}_2^R = R^{(2)} (R^{(2)}+1)$, energy spectrum corresponding to the Hamiltonian (\ref{eq:Hamgr42}) is given as
\be
E = \frac{1}{ 2 M \ell^2} \left( {\mathcal C}_2(p,q,r) - R^{(2)} (R^{(2)}+1) - \frac{n^2}{2}  \right ) \,.
\ee
This can be rewritten in terms of $q_2\,,n\,,p$ using (\ref{eq:C2}), (\ref{eq:chgn}), assuming $ p > r $ and introducing $m$ via $r = p - 2m$ ($m=0 \,, \cdots\,,\frac{p}{2}$ if $p$ is even and $m=0 \,, \cdots \,, \frac{p-1}{2}$ if $p$ is odd) as
\begin{align}
E&=\frac{1}{ 2 M \ell^2}\left (2q_2^2+2q_2(n+p+2)+n(p+2)+p^2+3p+m^2-m(p+1)-R^{(2)} (R^{(2)}+1)\right) \,.
\end{align}
In order to obtain the lowest energy we have to take the maximum value of the spin $R^{(2)}_{max} = \frac{r+p}{2}=p-m$.  Then, the energy spectrum becomes
\begin{equation}
E = \frac{1}{ 2 M \ell^2}\left (2q_2^2+2q_2(n+R^{(2)}+m+2)+n(R^{(2)}+m+2)+(R^{(2)}+m)(2+m)\right) \,.
\end{equation} 
The LLL energy at fixed background fields $R^{(2)}$ and $n$, is obtained for $q_2=m=0$ as follows:
\begin{equation}
E_{LLL} = \frac{1}{2 M \ell^2} \left(n(R^{(2)}+2)+2R^{(2)}\right) \,.
\label{eq:lenergy}
\end{equation}

The wave functions in the present case can be written in the form
\be
\mathcal{D}^{(p\,, q_1 + q_2\,, r)}_{L^{(1)}L^{(1)}_{3}L^{(2)}L^{(2)}_{3} L_{15}\, ; \, 0\,,0\,, R^{(2)} \,, R^{(2)}_3 \,, \frac{n}{\sqrt{2}}}(g) \,,
\label{eq:wf1}
\ee
where $R^{(2)}$ is given in (\ref{eq:spins1}). 

In order to have finite energy eigenvalues in the thermodynamic limit $\ell \rightarrow \infty \,, {\cal N} \rightarrow \infty$, the scales of $n$ and $R^{(2)}$ in terms of the powers of $\ell$ have to be determined. For a pure $SU(2)$ background ($n=0 \,, R^{(1)}= 0 \,, R^{(2)}\neq 0)$, $R^{(2)}$ should scale in the thermodynamic limit as $R^{(2)}\sim \ell^2$. The number of fermions in the LLL with $\nu =1$ is
\begin{equation}
{\cal N}=\mbox{dim}(R^{(2)}\,, 0\,, R^{(2)})=\frac{1}{12}(R^{(2)}+2)^2(2R^{(2)}+3)(R^{(2)}+1)^2 \underset{R^{(2)} \rightarrow \infty}{\longrightarrow} \frac{\left(R^{(2)}\right)^5}{6}
\end{equation}
and the corresponding spatial density is
\begin{equation}
\rho \sim \frac{{\cal N}} {\frac{\pi^4 \ell^8}{12}(2R^{(2)}+1)} \underset{\ell \rightarrow \infty \,, {\cal N} \rightarrow \infty}{\longrightarrow} \frac{\left (R^{(2)} \right )^4}{\pi^4\ell^8},
\end{equation}
which is finite.

When both $U(1)$ and $SU(2)$ backgrounds are present (i.e. $n\neq0,R^{(1)} =0 \,, R^{(2)} \neq 0$), just like the case of 
${\mathbb C}P^2$ reviewed in the previous section, we may choose either one of $n$ or $R^{(2)}$ to scale like $\ell^2$. Taking $n \sim \ell^2$ and $R^{(2)}$ to be finite in thermodynamic limit, we again get a finite spatial density
\begin{equation}
\rho \sim  \frac{{\cal N}} {\frac{\pi^4 \ell^8}{12}(2R^{(2)}+1)} \underset{\ell \rightarrow \infty \,, {\cal N}\rightarrow \infty}{\longrightarrow} \frac{n^4}{2 \pi^4\ell^8 R^{(2)}} \,,
\end{equation}
where we have the number of fermions ${\cal N}$ in the LLL with $\nu =1$ given in this case as
\begin{equation}
\mbox{dim}(R^{(2)}\,, n\,, R^{(2)})=\frac{1}{12}(R^{(2)}+n+2)^2(2R^{(2)}+n+3)(R^{(2)}+1)^2 (n+1) \underset{n \rightarrow \infty \,, R^{(2)}  \rightarrow \mbox{finite}}{\longrightarrow} \frac{n^4}{12} \,.
\end{equation} 

Before closing this subsection, we note that interchanging the Young tableaux of two $SU(2)$'s amounts to interchanging $R^{(1)}$ and $R^{(2)}$ in (\ref{eq:spins1}), and also a flip in the sign of the $U(1)$ charge. In the relevant formulas above, one can compensate for these changes by replacing $R^{(2)}$ with $R^{(1)}$ and substituting $|n|$ for $n$.

\subsection{$SU(2) \times SU(2)$ gauge field background}

Now we need to restrict ${\cal D}^{(p,q,r)}$ to those wave functions that transform as an IRR $(R^{(1)} \,, R^{(2)})$ of $SU(2) \times SU(2)$, and carry a $U(1)$ charge. It is useful to partition IRR of $SU(4)$ as $(p_1+p_2 \,, q_1+q_2 + x \,, r)$. There are now two classes of branchings which differ in their $U(1)$ charge as given in terms of $p_1,p_2,q_1,q_2$ and $r$ below.

If $q_2=0$, the branching with maximal $R^{(2)}$ value is
\begin{align}
 \overbrace{\ray{\yng(2,2,2)}\rar{$~\cdots$}}^r\overbrace{\ra{\yng(2,2)}\rar{$~\cdots$}}^q\overbrace{\rar{\yng(2)}
\rar{$~\cdots$}}^p&\rar{$\longrightarrow~$} \overbrace{\ra{\yng(2,2)}\rar{$~\cdots$}}^{r}\overbrace{\ra{\yng(1,1)}\rar{$~\cdots$}}^{q_1}\overbrace{{\rar{\yng(1)}\rar{$~\cdots$}}}^x\overbrace{{\rar{\yng(1)}\rar{$~\cdots$}}}^{p_1}\rar{$\otimes~$}\overbrace{\rar{\yng(1)}\rar{$~\cdots$}}^{r}\overbrace{{\rar{\yng(1)}\rar{$~\cdots$}}}^{p_2}\overbrace{{\rar{\yng(1)}\rar{$~\cdots$}}}^{x}
\end{align}
As $R^{(2)}$ decreases down from its maximal value $R^{(2)} = \frac{r +p_2 +x}{2}$ in increments of $1§$, the total number of boxes in each $SU(2)$ does not vary so we have, with $q=q_1+x$, 
\beq
n = \frac{1}{2}(2q_1-(p_2-p_1-r)) \,.
\label{eq:chg2}
\eeq

Suppose now that $q_2\neq 0$. This may happen only if all $p$-boxes are already in the tableaux of the second $SU(2)$ in the branching; thus we must have that $p_1 =0$. Once again, we have the branching with the maximal $R^{(2)}$ value as
\begin{align}
 \overbrace{\ray{\yng(2,2,2)}\rar{$~\cdots$}}^r\overbrace{\ra{\yng(2,2)}\rar{$~\cdots$}}^q\overbrace{\rar{\yng(2)}
\rar{$~\cdots$}}^p&\rar{$\longrightarrow~$} \overbrace{\ra{\yng(2,2)}\rar{$~\cdots$}}^{r}\overbrace{\ra{\yng(1,1)}\rar{$~\cdots$}}^{q_1}\overbrace{{\rar{\yng(1)}\rar{$~\cdots$}}}^x\rar{$\otimes~$}\overbrace{\ra{\yng(1,1)}\rar{$~\cdots$}}^{q_2}\overbrace{{\rar{\yng(1)}\rar{$~\cdots$}}}^{p}\overbrace{{\rar{\yng(1)}\rar{$~\cdots$}}}^{r}\overbrace{{\rar{\yng(1)}\rar{$~\cdots$}}}^{x}
\end{align}
and the $U(1)$ charge is now (with $p=p_2$)
\beq
n = \frac{1}{2} \left (2(q_1-q_2) - (p_2-r) \right) \,.
\label{eq:chg3}
\eeq

Using both of the tableaux we observe that the first $SU(2)$ in the branching takes the value
\begin{equation}
R^{(1)}=\frac{p_1+x}{2} \,, \quad 0 \leq x \leq q \,, \quad 0 \leq p_1 \leq p .
\label{eq:R1}
\end{equation}
For this value of $R^{(1)}$ the second $SU(2)$ takes on values between $R^{(2)}_{max} = \frac{S}{2}$ and $R^{(2)}_{min} = \frac{|2 {\cal M} -S |}{2} $ 
\begin{equation}
\frac{| 2 {\cal M}- S |}{2}  \leq R^{(2)} \leq \frac{S}{2} \,, \quad S=p_2+x+r \,, 
\label{eq:spin2}
\end{equation} 
where ${\cal M}$ is defined as the largest among the integers $p_2 \,, x$ and $r$.

We consider the cases $q_2 =0$ and $q_2 \neq 0$ with the $U(1)$ charges given in (\ref{eq:chg2}) and (\ref{eq:chg3})
separately to determine the energy spectrum corresponding to the Hamiltonian (\ref{eq:HamgrN2}). We have that 
\begin{equation}
E = \frac{1}{ 2M \ell^2} \left({\mathcal C}_2(p,q,r) - R^{(1)} (R^{(1)}+1)- R^{(2)} (R^{(2)}+1)- \frac{n^2}{2} \right) \,.
\label{eq:E22}
\end{equation}
For the case $q_2=0$, we have the condition that
\begin{equation}
m := \frac{p_2-p_1-r}{2}
\label{eq:m}
\end{equation}
is an integer to ensure that $n$ is so. Let us assume that $p_2 > p_1+r$ so that $m$ is positive. 

In order to obtain the lowest energy eigenvalues we use (\ref{eq:R1}) together with the maximum value of $R^{(2)}$ as given in (\ref{eq:spin2}).  Next, we eliminate $p_2 \,,q_1 \,, x$ and $r$ in favor of $n \,, R^{(1)} \,, R^{(2)} \,, p_1,$ and $m$ (explicitly we have $p_2 = R^{(2)} - R^{(1)} + p_1 + m$, $q_1 = n +m$, $x = 2 R^{(1)} - p_1$ and $r = R^{(2)} - R^{(1)} - 2 m$) to get
\begin{multline}
E = \frac{1}{2M \ell^2} \bigg({\mathcal C}_2 \left ( R^{(2)}-R^{(1)} + 2 p_1 + m \,, n+m+2R^{(1)}-p_1\,, R^{(2)}-R^{(1)}-m \right) \\ 
\hskip 9cm - R^{(1)} (R^{(1)}+1) - R^{(2)} (R^{(2)}+1) - \frac{n^2}{2} \bigg) \\
= \frac{1}{2M \ell^2} \big ( p_1^2+p_1(m+R^{(2)}-R^{(1)}+1)+m^2+m(R^{(1)}+R^{(2)}+n+2) \\
+n(R^{(1)}+R^{(2)}+2)+2R^{(2)} \big) \,,
\label{eq:levels1}
\end{multline}
where $R^{(2)} > R^{(1)}$ due to the assumption $p_2 > p_1+r$. For fixed $R^{(1)},R^{(2)}$, and $n$ Landau levels are controlled by the two integers $p_1$ and $m$. Taking $p_1=m=0$ results in the LLL energy 
\begin{equation}
E_{LLL} = \frac{1}{2M \ell^2} \left (n(R^{(1)}+R^{(2)}+2)+2R^{(2)} \right) \,.
\label{eq:eneigen}
\end{equation}
We note that assuming $p_2 < p_1+r$ flips the sign of $m$ and in (\ref{eq:levels1}) $m \rightarrow - m.$\footnote{The energy levels are still, of course, positive as can easily be checked.} 

It is also important to remark that for $R^{(1)} = R^{(2)} =R$, we have $p_1 = p_2 + r$ and thus
\be
{\tilde m} : = \frac{p_1 + r - p_2}{2} = r \,,
\ee
and the energy levels are given by
\beqa
E &=& \frac{1}{2M \ell^2} \left({\mathcal C}_2 \left ( 2 p_1 - r \,, n-r + 2R - p_1 \,, r \right) - 2 R (R +1) - \frac{n^2}{2} \right) \nn \\
&=& \frac{1}{2M \ell^2} \left ( 2 R + p_1 (1 + p_1 - {\tilde m}) + (n - {\tilde m}) (2 + 2 R - {\tilde m}) \right ) \,.
\eeqa
The energy values here are positive since $p_1 \geq {\tilde m}$, $n \geq {\tilde m}$, and $2 R - {\tilde m} \geq 0$ by construction.
The LLL energy is given by $p_1 = {\tilde m} = 0$, which is indeed the same as the one obtained from (\ref{eq:eneigen}) when $R:=R^{(1)} = R^{(2)} $.

The case $p_1=0$ may be treated along similar lines. We have that 
\begin{equation}
m: = \frac{p-r}{2}
\end{equation}
is an integer for the same reason that $n$ is so. Let us assume $p > r$ so that $m$ is positive. In this case we can write $p \,,q_1 \,, x$ and $r$ in terms of $n \,, R^{(1)} \,, R^{(2)} \,, q_2$ and $m$. Hence we find for the lowest energy eigenvalues 
\begin{multline}
E = \frac{1}{2M \ell^2} \bigg({\mathcal C}_2 \left ( R^{(2)}-R^{(1)} + m \,, 2q_2 +2R^{(1)} + n+m\,, R^{(2)}-R^{(1)}-m \right) \\ 
\hskip 9cm - R^{(1)} (R^{(1)}+1) - R^{(2)} (R^{(2)}+1) - \frac{n^2}{2} \bigg) \\
= \frac{1}{2M \ell^2} \big (2q_2^2+2q_2(n+R^{(1)}+R^{(2)}+m+2)+n(R^{(1)}+R^{(2)}+2)+m^{ 2} \\ 
+m(R^{(1)}+R^{(2)}+n+2)+2R^{(2)} \big).
\label{eq:en2}
\end{multline}

We note that here we do have the condition $R^{(2)} > R^{(1)}$ as well. In this case $q_2$ and $m$ specify the Landau levels. We take $q_2=m=0$ in (\ref{eq:en2}) to obtain the LLL energy and this yields the same result given in (\ref{eq:eneigen}) as expected. 

LLL energy for $R^{(2)}<R^{(1)}$ can be found by interchanging $R^{(1)}$ and $R^{(2)}$ in (\ref{eq:eneigen}) and taking $n$ to $-n$ where now $n < 0$. This gives
\begin{equation}
E_{LLL}=\frac{1}{2 M \ell^2} \left (- n(R^{(2)}+R^{(1)}+2)+2R^{(1)}\right).
\end{equation}

We do have two distinct cases to consider in the thermodynamic limit.  For a pure $SU(2) \times SU(2)$ background: $n=0, R^{(1)} \neq 0,R^{(2)}\neq 0$, both $R^{(1)}$ and $R^{(2)}$ should scale in the thermodynamic limit as $\ell^2$. The number of fermions in the LLL with $\nu =1$ is 
\begin{equation}
\mbox{dim}(R^{(2)}-R^{(1)}\,, 2R^{(1)}\,, R^{(2)}-R^{(1)}) \sim 4{R^{(1)}}^5R^{(2)} \,,
\end{equation}
and the corresponding spatial density in this limit is 
\begin{equation}
\rho \sim \frac{4 R^{(1)^{5}}R^{(2)}}{\pi^4\ell^8(2R^{(1)}+1)(2R^{(2)}+1)} \underset{\ell \rightarrow \infty \,, {\cal N} \rightarrow \infty}{\longrightarrow} \mbox{finite} \,.
\end{equation}
For the nonzero background $n\neq 0,R^{(1)}\neq0,R^{(2)}\neq 0$ we have three parameters $n$, $R^{(1)}$ and $R^{(2)}$. We can choose, say $n$ to scale like $\ell^2$ and the others to remain finite in thermodynamic limit. For $\nu =1$ we get
\begin{equation}
\mbox{dim}(R^{(2)}-R^{(1)}\,, 2R^{(1)}+n\,, R^{(2)}-R^{(1)}) {\longrightarrow}  \, \, n^4 \,
\end{equation} and the spatial density is
\begin{equation}
 \rho \sim \frac{n^4}{\pi^4\ell^8(2R^{(1)}+1)(2R^{(2)}+1)}\longrightarrow \mbox{finite}.
\end{equation}

\section{Landau Problem on $ \mathbf{Gr}_2(\mathbb{C}^N)$}

We are now ready to generalize the results of the previous section to all Grassmannians $\mathbf{Gr}_2(\mathbb{C}^N)$. It is useful to write down the coset realization 
\begin{equation} \mathbf{Gr}_2(\mathbb{C}^N)= \frac{SU(N)}{S[U(N-2)\times U(2)]} \sim \frac{SU(N)}{SU(N-2)\times SU(2) \times U(1)} \,. 
\label{eq:grn2}
\end{equation} 

The $SU(N)$ Wigner $\mathcal{D}$-functions for $g \in SU(N)$,
\begin{equation}
{\mathcal D}^{(P_1,P_2,P_3,\dots,P_{N-2},P_{N-1})}_{L^{SU(N-2)}\,, L\,, L_3\,, L_{N^2-1}\,, R^{SU(N-2)}\,, R\,, R_3\,, R_{N^2-1}} (g) \,,
\label{eq:sunDD}
\end{equation}
carrying the IRR $(P_1\,, P_2\,,P_3\,,\cdots\,, P_{N-2} \,, P_{N-1})$ labelled by $N-1$ non-negative integers, may be appropriately restricted to obtain the harmonics and local sections of bundles over $\mathbf{Gr}_2(\mathbb{C}^N)$. Let us denote the left- and the right-invariant vector fields on $SU(N)$ by $L_\alpha$ and $R_\alpha$ ($\alpha: 1\,, \cdots \,, N^2-1$); they satisfy the Lie algebra commutation relations for $SU(N)$. In (\ref{eq:sunDD}), $L^{SU(N-2)}$ and $R^{SU(N-2)}$ stand for the suitable sets of left and right quantum numbers, which we will not need in what follows.

The real dimension of $\mathbf{Gr}_2(\mathbb{C}^N)$ is $4N-8$ and tangents along $\mathbf{Gr}_2(\mathbb{C}^N)$ may be parametrized by the $4N-8$ right-invariant fields, $R_\alpha$, ($\alpha:N^2-4N+7, \cdots,N^2-2$). Consequently, the Hamiltonian may be written as
\beqa
H &=& \frac{1}{2 m\ell^2} \sum^{N^2-2}_{\alpha = N^2-4N+7} R_\alpha^2 \nn \\
&=&\frac{1}{2 m\ell^2} \left (\mathcal{C}_2^{SU(N)} - {\mathcal C}_2^{SU(N-2)} - {\mathcal C}_2^{SU(2)} - R_{N^2-1}^2 \right) \,.
\label{eq:HamgrN2}
\eeqa
Here for future use we give the eigenvalue of $\mathcal{C}_2^{SU(N)}$ in the IRR $(P_1\,, P_2\,, 0 \,,\cdots\,, 0 \,, $ $P_{N-2} \,, P_{N-1})$, which reads
\begin{eqnarray}
\mathcal{C}_2(P_1,P_2,0,\dots,0,P_{N-2},P_{N-1}) & = &(\frac{N-1}{2N})P_1^2+(\frac{N-2}{N})P_2^2+(\frac{N-2}{N})P_{N-2}^2\nonumber \\
&+&(\frac{N-1}{2N})P_{N-1}^2
+(\frac{N-2}{N})P_1P_2+\frac{2}{N}P_1P_{N-2}\nonumber\\&+&\frac{1}{N}P_1P_{N-1}+\frac{4}{N}P_2P_{N-2}+\frac{2}{N}P_2P_{N-1}
\nonumber\\&+&(\frac{N-2}{N})P_{N-2}P_{N-1}+(\frac{N-1}{2})P_1+(N-2)P_2 
\nonumber\\&+&(N-2)P_{N-2}
+(\frac{N-1}{2})P_{N-1},
\label{eq:casuN}
\end{eqnarray}
and the dimension of this representation is given in the appendix B.

In order to obtain the wave functions with only a $U(1)$ background gauge field, we consider those ${\cal D}$-functions that transform trivially under the right action of $SU(N-2)$ and  $SU(2)$, and carry a right $U(1)$ charge. This means these wave functions remain singlets under $SU(N-2)$ and $ SU(2)$ with non-zero ${\mathcal C}_2^{SU(N-2)} ,{\mathcal C}_2^{SU(2)} $ eigenvalues and a non-zero $R_{N^2-1}$ eigenvalue. 

The branching $SU(N) \supset SU(N-2)\times SU(2)\times U(1)$ may be utilized for this purpose. In order to have both $SU(N-2)$ and $ SU(2)$ as singlets in the branching we must require all $P_i$ except $P_1 \,, P_2 \,, P_{N-2} \,, P_{N-1}$ to vanish, and also $P_{N-1} = P_1$. In terms of Young tableaux, this branching can be shown by
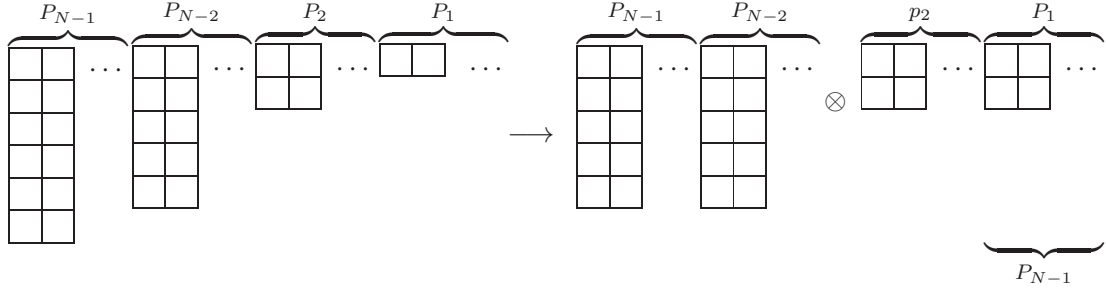
\begin{align*}
 \overbrace{\ray{\yng(2,2,2,2,2,2)\rarr{\rar{$~\cdots$}}}}^{P_{N-1}}\overbrace{\ra{\yng(2,2,2,2,2)}\rar{\rarr{$~\cdots$}}}^{P_{N-2}}\overbrace{\rar{\rar{\yng(2,2)}\rarr{$~\cdots$}}}^{P_2}\overbrace{\rarr{\rar{\yng(2)}
\rar{$~\cdots$}}}^{P_1}&\rarr{$\longrightarrow~$}
\overbrace{\ra{\yng(2,2,2,2,2)}\rarr{\rar{$~\cdots$}}}^{P_{N-1}}\overbrace{\ra{\yng(2,2,2,2,2)}\rarr{\rar{$~\cdots$}}}^{{P_{N-2}}}\rar{\rar{$\otimes~$}}\overbrace{\rar{\rar{\yng(2,2)}}\rar{\rarr{$~\cdots$}}}^{p_{2}}\overbrace{{\underbrace{\rar{\rar{\yng(2,2)}}\rar{\rarr{$~\cdots$}}}_{P_{N-1}}}}^{P_1}
\end{align*}
where the tableaux on the l.h.s represent the IRR $(P_1\,, P_2\,, 0\,,\cdots\,,0\,, P_{N-2} \,, P_1)$ of $SU(N)$. The tableaux on the r.h.s are those of $SU(N-2)$ and $SU(2)$, respectively, and both are singlets in this case.

From (\ref{eq:charge}) we compute the $U(1)$ charge as
\begin{eqnarray}
n&=&\frac{1}{2(N-2)}(2J_1-(N-2)J_{2}) \nn \\
&=& P_{N-2}-P_2 \,.
\label{SU(N)charge}
\end{eqnarray} 
The relation between eigenvalues of $R_{N^2-1}$ and $n$ is found to be (see Appendix A)
\begin{equation}
R_{N^2-1}= -\sqrt{1-\frac{2}{N}} n 
\label{R_matrix} \,.
\end{equation}

The energy spectrum of the Hamiltonian is 
\beqa
E &=& \frac{1}{2 M \ell^2} \left (\mathcal{C}_2^{SU(N)} -  (1-\frac{2}{N}  ) n^2 \right ) \\
&=& \frac{1}{2 M \ell^2} \left (P_1^2 + (2-\frac{4}{N})P_2^2+(N-1+2n)P_1+2(n+N-2+\frac{2}{N})P_2+4P_1P_2+n(N-2) \right ) \nn \,,
\eeqa
where we have used (\ref{eq:casuN}) with $P_{N-1} = P_1$ and $P_{N-2}= P_2 +n$. The integers $P_1$ and $P_2$ are in fact considered to be the Landau level indices. The LLL energy can be obtained by setting $P_1=P_2=0$, which is
\begin{equation}
E_{LLL}=\frac{Nn-2n}{2 M \ell^2} \,.
\end{equation}
The corresponding wave functions may be expressed by
\be
{\mathcal D}^{(P_1\,, P_2 \,, 0 \,, \cdots\,, 0\,, P_{n-2} = P_2 +n\,, P_{n-1} = P_1)}_{L^{SU(N-2)}\,, L\,, L_3\,, L_{N^2-1}\,, 0 \,, 0 \,, 0 \,, -\sqrt{1-\frac{2}{N}} n} (g) \,.
\ee

Spatial density of particles in the thermodynamic limit is computed in a manner analogous to those for the case of $\mathbf{Gr}_2(\mathbb{C}^4)$. We have ${\text {vol}}(\mathbf{Gr}_2(\mathbb{C}^N))=\frac{\pi^{2(N-2)}}{(N-2)!(N-1)!} \ell^{4N-8}$ from \eqref{volume} and in the LLL, $P_{N-2}=n$, $P_1=P_2=P_{N-1}=0$, with $\nu=1$. Correspondingly, the dimension formula \eqref{eq:dimsuN} for the LLL with $\nu=1$ reduces to 
\begin{equation}
{\cal N} = \dim(0\,,0\,,\cdots\,,n\,,0) = \frac{(n+N-3)!(n+N-4)!(n+N-2)^2(n+N-1)(n+N-3)}{(N+1)!(N-2)!n!(n+1)!}\,.
\end{equation}
In the thermodynamic limit ($\ell\rightarrow \infty$ and ${\cal N} \rightarrow \infty$), the density of the states takes the form
\begin{equation}
\rho=\frac{{\cal N}}{\frac{\pi^{2(N-2)}}{(N-2)!(N-1)!} \ell^{4N-8}} \longrightarrow \frac{n^{2N-4}}{\ell^{4N-8}}=\left(\frac{B}{2\pi}\right)^{2N-4} \,.
\end{equation}

For the case of both $SU(2)$ and $U(1)$ background gauge fields, the spectrum of the Hamiltonian and the wave functions are obtained in a similar manner. We still have to demand all $P_i$ except $P_1 \,, P_2 \,, P_{N-2} \,, P_{N-1}$ to vanish, but no longer impose the condition $P_{N-1} = P_1$. The relevant branching of $SU(N)$ is now given by the Young tableaux below:
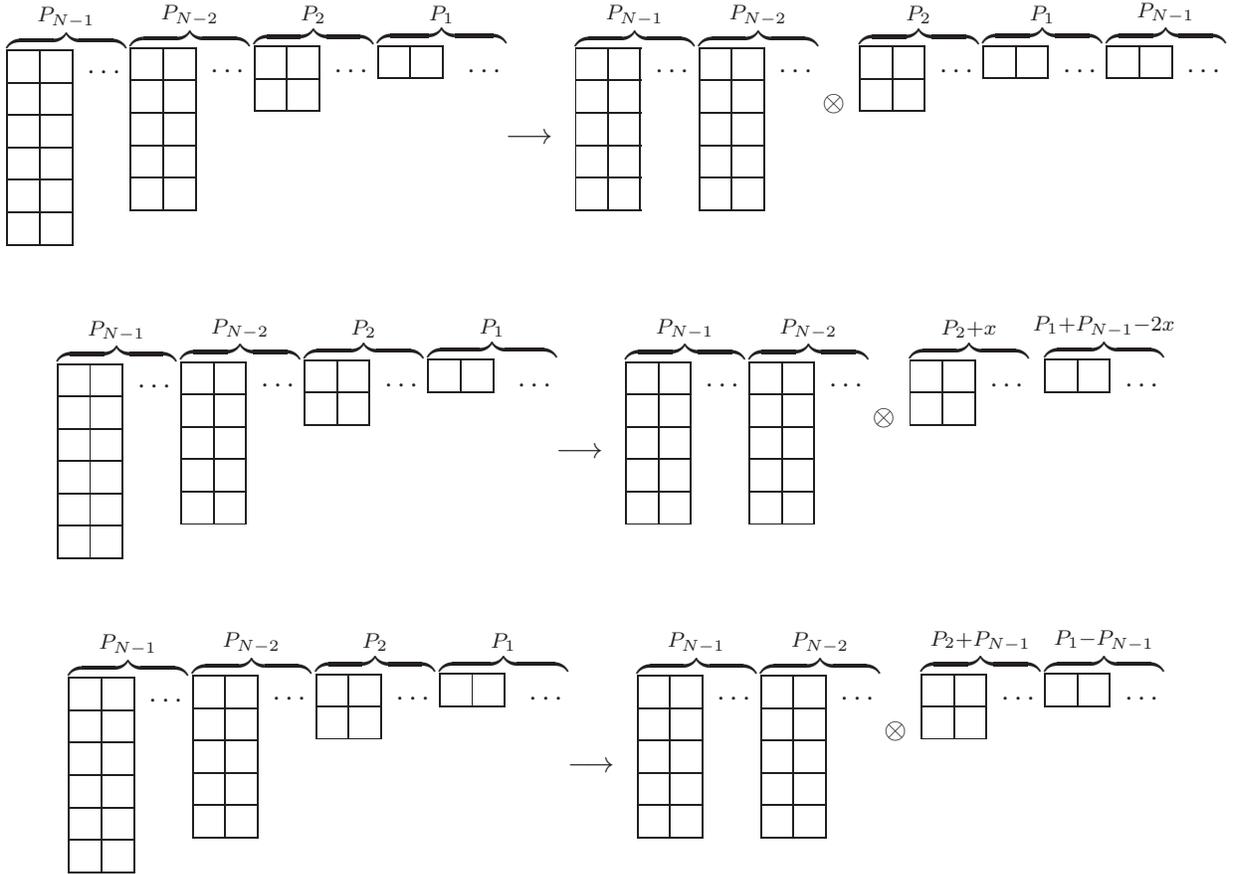
\begin{align*}
 \overbrace{\ray{\yng(2,2,2,2,2,2)\rarr{\rar{$~\cdots$}}}}^{P_{N-1}}\overbrace{\ra{\yng(2,2,2,2,2)}\rar{\rarr{$~\cdots$}}}^{P_{N-2}}\overbrace{\rar{\rar{\yng(2,2)}\rarr{$~\cdots$}}}^{P_2}\overbrace{\rarr{\rar{\yng(2)}
\rar{$~\cdots$}}}^{P_1}&\rarr{$\longrightarrow~$}
\overbrace{\ra{\yng(2,2,2,2,2)}\rarr{\rar{$~\cdots$}}}^{P_{N-1}}\overbrace{\ra{\yng(2,2,2,2,2)}\rarr{\rar{$~\cdots$}}}^{{P_{N-2}}}\rar{\rar{$\otimes~$}}\overbrace{\rar{\rar{\yng(2,2)}}\rar{\rarr{$~\cdots$}}}^{{P_{2}}}\overbrace{{\rar{\rarr{\yng(2)}}\rar{\rarr{$~\cdots$}}}}^{{P_{1}}}\overbrace{{\rar{\rarr{\yng(2)}}\rar{\rarr{$~\cdots$}}}}^{{P_{N-1}}}
\end{align*}
\begin{align*}
\overbrace{\ray{\yng(2,2,2,2,2,2)\rarr{\rar{$~\cdots$}}}}^{P_{N-1}}\overbrace{\ra{\yng(2,2,2,2,2)}\rar{\rarr{$~\cdots$}}}^{P_{N-2}}\overbrace{\rar{\rar{\yng(2,2)}\rarr{$~\cdots$}}}^{P_2}\overbrace{\rarr{\rar{\yng(2)}
\rar{$~\cdots$}}}^{P_1}&\rarr{$\longrightarrow~$}
\overbrace{\ra{\yng(2,2,2,2,2)}\rarr{\rar{$~\cdots$}}}^{P_{N-1}}\overbrace{\ra{\yng(2,2,2,2,2)}\rarr{\rar{$~\cdots$}}}^{{P_{N-2}}}\rar{\rar{$\otimes~$}}\overbrace{\rar{\rar{\yng(2,2)}}\rar{\rarr{$~\cdots$}}}^{{P_{2}+x}}\overbrace{{\rar{\rarr{\yng(2)}}\rar{\rarr{$~\cdots$}}}}^{{{ P_1 + P_{N-1}-2x}}}
\end{align*}
\begin{align*}
\overbrace{\ray{\yng(2,2,2,2,2,2)\rarr{\rar{$~\cdots$}}}}^{P_{N-1}}\overbrace{\ra{\yng(2,2,2,2,2)}\rar{\rarr{$~\cdots$}}}^{P_{N-2}}\overbrace{\rar{\rar{\yng(2,2)}\rarr{$~\cdots$}}}^{P_2}\overbrace{\rarr{\rar{\yng(2)}
\rar{$~\cdots$}}}^{P_1}&\rarr{$\longrightarrow~$}
\overbrace{\ra{\yng(2,2,2,2,2)}\rarr{\rar{$~\cdots$}}}^{P_{N-1}}\overbrace{\ra{\yng(2,2,2,2,2)}\rarr{\rar{$~\cdots$}}}^{{P_{N-2}}}\rar{\rar{$\otimes~$}}\overbrace{\rar{\rar{\yng(2,2)}}\rar{\rarr{$~\cdots$}}}^{{P_{2}+P_{N-1}}}\overbrace{{\rar{\rarr{\yng(2)}}\rar{\rarr{$~\cdots$}}}}^{{ P_1-P_{N-1}}}
\end{align*}
where the branching rule for maximum, generic and minimum $SU(2)$ spin are given, respectively and $0 \leq x \leq P_{N-1}$. We have assumed that $P_1\ge P_{N-1}$. The $SU(2)$ spin interval is then
\be
R = \frac{P_1 - P_{N-1}}{2} \,,\cdots \,, \frac{{P_{N-1}+P_1}}{2} \,, 
\label{SU(2)charge}
\ee
and the $U(1)$ charge is given by
\begin{equation}
\label{eq:chargeSU(N) with R} n = \frac{1}{2} \left (P_{N-1}+2(P_{N-2}-P_2)-P_1 \right) \,.
\end{equation} 
By the Dirac quantization condition $n$ should be an integer so we must have that 
\begin{equation}
\label{eq:integer_suN} 
m: = \frac{P_1 - P_{N-1}}{2} \,,
\end{equation}
is an integer taking values within the interval $m=0,\cdots, \frac{P_1}{2}$ if $P_1$ is even and $m=0,\cdots, \frac{P_1-1}{2}$ if $P_1$ is odd.
The energy spectrum corresponding to the Hamiltonian (\ref{eq:HamgrN2}) reads
\begin{equation}
\label{eigenvalueSU(N) with R} E = \frac{1}{2 M \ell^2} \left (\mathcal{C}_2^{SU(N)} - R(R+1) -  (1-\frac{2}{N} ) n^2 \right) \,.
\end{equation}
This equation can be re-written in terms of $P_2, P_1,m$ and $n$ by using \ref{eigenvalueSU(N) with R}, \ref{eq:chargeSU(N) with R} and \ref{eq:integer_suN}:
\begin{eqnarray}
E & = & \frac{1}{2M\ell^2} \bigg ( (\frac{N-1}{2N})P_1^2+(\frac{N-2}{N})P_2^2+(\frac{N-2}{N})(n^2+m^2+2nm+P_2^2+2nP_2+2mP_2) \nonumber \\
&+& (\frac{N-1}{2N})(4m^2+P_1^2-4mP_1)
+(\frac{N-2}{N})P_1P_2+\frac{2}{N}P_1(n+m+P_2)-\frac{1}{N}(2mP_1-P_1^2)\nonumber\\&+&\frac{4}{N}P_2(n+m+P_2)-\frac{2}{N}P_2(2m-P_1)-(\frac{N-2}{N})(2m-P_1)(n+m+P_2)+(\frac{N-1}{2})P_1\nonumber\\&+&(N-2)P_2 
+(N-2)P_{N-2} +(\frac{N-1}{2})(-2m+P_1)-(\frac{N-2}{N})n^2-R(R+1) \bigg ) \,.
\end{eqnarray}
Taking the maximum value of the spin $R$,
\begin{equation}
R = \frac{P_{N-1}+P_1}{2}= P_1 - m \,,
\end{equation}
 the lowest energy becomes
\begin{eqnarray}
E &=& \frac{1}{2M\ell^2} \bigg ( (\frac{N-1}{2N})(2R^2+2m^2)+ \frac{N-2}{N}(2P_2^2+mn+2nP_2+2RP_2+2mP_2+Rn+Rm)\nonumber\\ &+&\frac{1}{N}(2Rn+4RP_2+2mn+R^2+m^2+2Rm+4P_2n+4P_2m+4P_2^2) \nonumber \\&+&(\frac{N-1}{2})(2R)+(N-2)(2P_2+n+m)-R(R+1) \bigg ) \,.
\end{eqnarray}
Once again, the LLL at fixed background charges $n$ and $R$ are controlled by two integers, $m$ and $P_2$. The LLL is  found by putting $P_2 = m = 0$. This gives the energy eigenvalue
\begin{equation}
E_{LLL}=\frac{1}{2 M \ell^2} \left (n R + (N - 2) (n + R) \right) \,,
\end{equation}
which collapses to (\ref{eq:lenergy}) for $N=4$ as expected. More generally, to match the formulas of this section to those for $N=4$, we note that the correspondence for the IRR labels is determined to be
\be
(p\,, q = q_1+q_2 \,,r) \longrightarrow (P_1\,, P_{2} = q_2 \,, 0\,,\cdots\,, P_{N-2} = q_1 \,, P_{N-1}) \,,
\ee

For a pure $SU(2)$ background $n=0,R\neq 0$, $R$ should scale in the thermodynamic limit as $R^{(2)}\sim \ell^2$. The number of fermions in the LLL with $\nu =1$ is ${\cal N} = \dim(R,0,\cdots,0,R)$ where
\begin{eqnarray}
\dim(R,0,\cdots,0,R) & = & \frac{1}{(N-1)!(N-2)!(N-3)!(R+1)!R! } ((R+N-3)!(N-4)!(R+N-3)!\nonumber\\&\times&(R+N-2) (R+1)(2R+N-1)(N-3)(R+N-2) ), \nonumber 
\end{eqnarray}
and the corresponding spatial density is
\begin{equation}
\rho \sim \frac{\cal{N}} {\ell^{4N-8}(2R+1)} \longrightarrow \frac{R^{2N-3}}{k \ell^{4N-8}(2R+1)} \longrightarrow \mbox{finite} \,.
\end{equation}

For both $U(1)$ and $SU(2)$ backgrounds $n\neq0,R\neq 0$, we can choose the scaling $n\sim \ell^2$ and keep $R$ finite
in thermodynamic limit. The ${\cal N}$ in the LLL with $\nu =1$ is
\be
{\cal N} = \dim(R \,, 0\,,\cdots\,, n\,, R) \longrightarrow n^{2N-4} \,,
\ee
and the spatial density reads
\begin{equation}
\rho \sim \frac{\cal{N}} {\ell^{4N-8}(2R+1)} \longrightarrow \frac{n^{2N-4}}{k \ell^{4N-8}(2R+1)} \longrightarrow \mbox{finite} \,.
\end{equation}

Before ending this section, let us briefly list a few of the results of our analysis for the Landau problem on $\mathbf{Gr}_2(\mathbb{C}^5)$. Labeling the IRR of $SU(5)$ with $(p\,,q\,,r\,,s)$, we find that the energy spectrum due to only an abelian monopole background is
\beqa
 E &=& \frac{1}{2M\ell^2} \left ( \mathcal{C}_2^{SU(5)} - \frac{3}{5}n^2 \right ) \nn \\ 
&=&{ \frac{1}{2M\ell^2} \left ( p^2+2q^2+2nq+2qp+pn+4p+6q+3n \right ) \,,}
\eeqa
where we have used $p=s$ and $r = n + q$ in $\mathcal{C}_2^{SU(5)}$. The numbers $p$ and $q$ play the role of Landau level indices. So the ground state energy is obtained by letting $p=q=0$, which yields
\be
E_{LLL}=\frac{3n}{2M\ell^2} \,,
\ee
and wave functions take the form
\be
{\mathcal D}^{(p\,, q \,, n+q \,, p)}_{L^{SU(3)}\,, L\,, L_3\,, L_{24}\,, 0 \,, 0 \,, 0 \,, -\sqrt{\frac{3}{5}} n} (g) \,.
\ee
With reference to \eqref{eq:dimsuN} the dimension of the $(0\,,0\,,n\,,0)$ representation gives the degeneracy of the LLL as follows:
\begin{equation}
\dim(0\,,0\,,n\,,0) = \frac{(n+2)!(n+1)!(n+3)^2(n+4)(n+2)}{4!3!n!(n+1)!}.
\end{equation}
Finally, the spatial density of fermions is readily computed to be
\begin{equation}
\rho \longrightarrow \frac{n^6}{\ell^{12}} = \left(\frac{B}{2\pi}\right)^6 \,.
\end{equation}

For $SU(2)$ and $U(1)$ backgrounds together, the energy spectrum reads
\be
E = \frac{1}{2M\ell^2} \left (\mathcal{C}_2^{SU(5)} - R(R+1) - \frac{3}{5}n^2 \right )  \,,
\ee
where $SU(2)$ has the spin range
\be
R = \frac{p-s}{2},\cdots,\frac{{s+p}}{2} \,,
\ee
assuming that $p > s$. The $U(1)$ charge now reads $n = \frac{1}{2} \left (s + 2(r - q)- p \right )$. Setting $ s = p - 2m$, the maximal $SU(2)$ charge $R = p - m$ gives the energy eigenvalues 
\begin{equation}
E = \frac{1}{2 M \ell^{2}}(m^{2}+2q^{2}+mn+2qn+2Rq+2mq+Rn+Rm+3R+6q+3n+3m) \,.
\end{equation}
Here applying the LLL condition gives the lowest energy as
\begin{equation}
E_{LLL}=\frac{1}{2 M \ell^{2}}( n(R+3) + 3R ) \,.
\end{equation}

\section{Local Form of the Wave Functions and the Gauge Fields}

In this section, we first provide the local form of the wave functions for solutions of the Landau problem on $\mathbf{Gr}_2(\mathbb{C}^4)$. For this purpose, we will utilize the well-known Pl\"ucker coordinates for $\mathbf{Gr}_2(\mathbb{C}^4)$. 

The Pl\"ucker coordinates for $\mathbf{Gr}_k(\mathbb{C}^N)$ are constructed out of a projective embedding, the so-called  Pl\"ucker embedding $\mathbf{Gr}_k(\mathbb{C}^n)\hookrightarrow \mathbf{P}\left(\bigwedge^k \mathbb{C}^n\right)$, which provides a one-to-one map between the set of $k$-dimensional subspaces of $\mathbb{C}^n$ (i.e. the Grassmannian $\mathbf{Gr}_k(\mathbb{C}^N)$) and a subset of the projective space of the $k^{th}$ exterior power of the vector space $\mathbb{C}^n$, where the latter is denoted as $\mathbf{P}\left(\bigwedge^k \mathbb{C}^n\right)$. This subset of $\mathbf{P}\left(\bigwedge^k \mathbb{C}^n\right)$ is a projective variety characterized by the intersection of quadrics induced by all possible relations between generalized Pl\"ucker coordinates. In what follows, we focus on the Pl\"ucker
embedding of $\mathbf{Gr}_2(\mathbb{C}^4)$; more details and  general discussions could be found in \cite{Ward-Wells, Harris}.

For $\mathbf{Gr}_2(\mathbb{C}^4)$ this construction entails the projective space $\mathbf{P} \left(\mathbb{C}^4 \wedge \mathbb{C}^4 \right) \equiv {\mathbb C}P^5$. Introducing two sets of complex coordinates $v_\alpha \,, w_\alpha$ $(\alpha =1\,, \cdots \,, 4)$, that is one set for each $\mathbb{C}^4$, a fully antisymmetric basis for the exterior product space $\mathbb{C}^4 \wedge \mathbb{C}^4$ would be given in the form of
\begin{equation}
P_{\alpha \beta} = \frac{1}{\sqrt{2}} (v_\alpha w_\beta - v_\beta w_\alpha ) \,.
\end{equation}
$P_{\alpha \beta}$ may be contemplated as the homogenous coordinates on ${\mathbb C}P^5$ with the identification $P_{\alpha \beta} \sim \lambda P_{\alpha \beta}$ where $\lambda \in U(1)$ and $\sum_{\alpha \,, \beta}^4|P_{\alpha \beta}|^2=1$.

The Pl\"ucker embedding of $\mathbf{Gr}_2(\mathbb{C}^4)$ in ${\mathbb C}P^5$ is given by the homogeneous condition 
\begin{equation}
\varepsilon_{\alpha \beta \gamma \delta} P_{\alpha \beta} P_{\gamma \delta} = P_{12} P_{34} - P_{13} P_{24} + P_{14}P_{23} = 0 \,,
\label{eq:pluckerrel}
\end{equation}
defining the Klein quadric $Q_4$ in ${\mathbb C}P^5$, which is complex analytically equivalent to $\mathbf{Gr}_2(\mathbb{C}^4)$. The homogeneous equation $\varepsilon_{\alpha \beta \gamma \delta} P_{\alpha \beta} P_{\gamma \delta} = 0$ is nothing but the restriction to a projective hypersurface of degree two, which is the quadric $Q_4$.

It is possible to employ $P_{\alpha \beta}$ to parametrize the columns of $g \in SU(4)$ in the IRR $(0,1,0)$; we choose a parametrization of the form\begin{equation} 
g:=\ \ \begin{blockarray}{rrrrrrrrrrrrr}
\begin{block}{(rrrrrrrrrrrrr)}
\BAmulticolumn{1}{c}{\multirow{6}{*}{$\vdots$}} & \vrule & \BAmulticolumn{1}{c}{\multirow{6}{*}{$\vdots$}} & \vrule & \BAmulticolumn{1}{c}{\multirow{6}{*}{$\vdots$}} & \vrule & \BAmulticolumn{1}{c}{\multirow{6}{*}{$\vdots$}} & \Vrule & P^{*}_{34} & P_{12}\\
& \vrule &  & \vrule & & \vrule & & \Vrule & - P^{*}_{24} & P_{13}\\
& \vrule &  & \vrule & & \vrule & & \Vrule & P^{*}_{23} & P_{14}\\
& \vrule &  & \vrule & & \vrule & & \Vrule & P^{*}_{14} & P_{23}\\
& \vrule &  & \vrule & & \vrule & & \Vrule & - P^{*}_{13} & P_{24}\\
& \vrule &  & \vrule & & \vrule & & \Vrule & P^{*}_{12} & P_{34}\\
\end{block}
\end{blockarray} \,,
\label{eq:su4par}
\end{equation}
where the orthogonality of the columns follow from the Pl\"ucker relation in (\ref{eq:pluckerrel}). For a short-hand notation, we will employ $g_{N 6} = P_N := P_{\alpha \beta}$, $g_{N 5} = \varepsilon_{NM} P_M^* = \varepsilon_{\alpha \beta \gamma \delta} P_{\gamma \delta}^*$ with $N \equiv \lbrack \alpha \beta \rbrack$, $N = 1 \,, \cdots \,, 6$ and $\alpha \beta = \left ( 12,13,14,23,24,34 \right )$.

The wave functions in the $U(1)$ background gauge field, $\mathcal{D}^{(0 \,, q_1 + q_2\,, 0)}_{L^{(1)}L^{(1)}_{3}L^{(2)}L^{(2)}_{3} L_{15}\, ; \, 0\,,0\,, 0\,, 0\,, \frac{n}{\sqrt{2}}}(g)$, are the sections of $U(1)$ bundle over $\mathbf{Gr}_2(\mathbb{C}^4)$, which fulfill the gauge transformation property 
\be
\mathcal{D}^{(0 \,, q_1 + q_2 \,, 0)} (g h) = \mathcal{D}^{(0 \,, q_1 + q_2 \,, 0)} (g e^{i \lambda_{15} \theta}) = \mathcal{D}^{(0 \,, q_1 + q_2 \,, 0)} (g) e^{i \frac{n}{\sqrt{2}} \theta} \,.
\label{eq:sections}
\ee
Using (\ref{eq:chargeU(1)}) for $\lambda_{15}$ and (\ref{eq:su4par}), this yields immediately 
\be
\mathcal{D}^{(0 \,, 1 \,, 0)} (g) \sim P_{\alpha \beta} \,.\label{oneparticlewavefunction}
\ee
We point out that the $(0,q,0)$ IRR is the $q$-fold symmetric tensor product of the $(0,1,0)$ representation; to wit, $(0,q,0) \equiv \prod_{\otimes q} (0,1,0)$. This can be shown by the symmetric tensor product $(\otimes_S)$ of $\tiny \yng(1,1)$ tableaux as
\begin{align*}
\yng(1,1)\ \otimes_S\ \yng(1,1)\ \otimes_S\ \cdots\ \otimes_S\   \yng(1,1)
\  \longrightarrow\ 
 \overbrace{\yng(3,3)\ \cdots\  \yng(1,1)}^{q}.
\end{align*}
We infer that
\be
\mathcal{D}^{(0 \,, q_1+q_2 \,, 0)} (g) \sim P_{\alpha_1 \beta_1} P_{\alpha_2 \beta_2} \cdots P_{\alpha_{q_1} \beta_{q_1}}  
P^*_{\gamma_1 \delta_1} P^*_{\gamma_2 \delta_2} \cdots P^*_{\gamma_{q_2} \delta_{q_2}} \,.
\label{eq:wavef}
\ee
So the LLL wave functions are those with $q_2=0$:
\be
\mathcal{D}^{(0 \,, q_1 \,, 0)}_{L^{(1)}L^{(1)}_{3}L^{(2)}L^{(2)}_{3} L_{15}\, ; \, 0\,,0\,, 0\,, 0\,, \frac{n}{\sqrt{2}}}(g) \sim P_{\alpha_1 \beta_1} P_{\alpha_2 \beta_2} \cdots P_{\alpha_{q_1} \beta_{q_1}} \,,
\label{eq:LLLwf}
\ee
which are holomorphic in the Pl\"ucker coordinates.

Another useful point to mention here is that, although the right-invariant vector fields on $SU(4)$ cannot be easily written down, the left-invariant vector fields can be easily given as \cite{Chris}
\beq
L_k = - v_{j}(\lambda_k)_{ij}\frac{\partial}{\partial v_{i}}-w_j(\lambda_k)_{ij}\frac{\partial}{\partial w_{i}}+v^{*}_{i}(\lambda_k)_{ij}\frac{\partial}{\partial v^{*}_{j}}+w^{*}_{i}(\lambda_k)_{ij}\frac{\partial}{\partial w^{*}_{j}}\,, \label{left-inv}
\eeq
where $\lambda_k$ $(k=1\,,\dots \,,15)$ are the Gell-Mann matrices for $SU(4)$. Choosing  complex vectors $\mathbf{v}$ and $\mathbf{w}$ to satisfy the orthonormality conditions 
\beq
v_i w^{*}_i = 0 \,, \quad  |\mathbf{v}|^2=|\mathbf{w}|^2=1 \,,
\label{orthono}
\eeq
and using the identity
\beq
\sum_{k=1}^{N^2-1}\lambda ^k_{ij}\lambda^k_{mn}= \frac{1}{2}\delta_{in}\delta_{jm}-\frac{1}{2N}\delta_{ij}\delta_{mn} \,,
\eeq
for $N=4$, the Casimir $\mathcal{C}_{2}^{SU(4)}$ may be realized as the differential operator:
\begin{align}
\mathcal{C}_{2}^{SU(4)}& = \frac{15}{8}\bigg(v_{i}\frac{\partial}{\partial v_{i}}+w_{i}\frac{\partial}{\partial w_{i}}+v_{i}^{*}\frac{\partial}{\partial v_{i}^{*}}+w_{i}^{*}\frac{\partial}{\partial w_{i}^{*}}\bigg)+\frac{3}{8}\bigg(v_{i}v_{j}\frac{\partial}{\partial v_{i}}\frac{\partial}{\partial v_{j}}+w_{i}w_{j}\frac{\partial}{\partial w_{i}}\frac{\partial}{\partial w_{j}}+\text{c.c.}\bigg)+\cr &-\frac{2}{8}\bigg(v_{i}w_{j}\frac{\partial}{\partial v_{i}}\frac{\partial}{\partial w_{j}}-v_{i}w_{j}^{*}\frac{\partial}{\partial v_{i}}\frac{\partial}{\partial w_{j}^{*}}+\text{c.c.}\bigg)+\frac{1}{8}\bigg(v_{i}v_{j}^{*}\frac{\partial}{\partial v_{i}}\frac{\partial}{\partial v_{j}^{*}}+w_{i}w_{j}^{*}\frac{\partial}{\partial w_{i}}\frac{\partial}{\partial w_{j}^{*}}+\text{c.c.}\bigg)+\cr &+v_{i}w_{j}\frac{\partial}{\partial v_{j}}\frac{\partial}{\partial w_{i}}+v_{i}^{*}w_{j}^{*}\frac{\partial}{\partial v_{j}^{*}}\frac{\partial}{\partial w_{i}^{*}}-\frac{\partial}{\partial v_{j}}\frac{\partial}{\partial v_{j}^{*}}-\
\frac{\partial}{\partial w_{j}}\frac{\partial}{\partial w_{j}^{*}} \,,
\label{casimir}
\end{align}
which clearly generates the eigenvalues $\frac{q^2}{2}+2q$ when applied to the wave functions (\ref{eq:wavef}).

The LLL with filling factor $\nu = 1$ has ${\cal N} = \mbox{dim}(0,1,0) = \frac{1}{12}(n+1)(n+2)^2(n+3)$ number of particles. Its multi-particle wave-function is given in terms of the Slater determinant as  
\beqa
\Psi_{MP} &=& \frac{1}{\sqrt{{\cal N}!}}\det\left(\begin{array}{cccc}
\Psi_{\Lambda_{1}}(P^{\mathbf{1}}) & \cdots &  & \Psi_{\Lambda_{\cal N}}(P^{\mathbf{1}})\\
\Psi_{\Lambda_{1}}(P^{\mathbf{2}}) & \cdots &  & \Psi_{\Lambda_{\cal N}}(P^{\mathbf{2}})\\
\vdots &  & \ddots & \vdots\\
\Psi_{\Lambda_{1}}(P^{\mathbf{{\cal N}}}) & \cdots &  & \Psi_{\Lambda_{{\cal N}}}(P^{\mathbf{{\cal N}}})
\end{array}\right) \nn \\
&=& \frac{1}{\sqrt{{\cal N}!}} \varepsilon^{\Lambda_1 \Lambda_2 \, \cdots \Lambda_n} \Psi_{\Lambda_1}(P^{(1)}) \Psi_{\Lambda_2} (P^{(2)}) \cdots \Psi_{\Lambda_N} (P^{(N)}) \,.\label{DetPsi}
\eeqa
Here $P^{\mathbf{i}}$ denotes the $i^{th}$ position fixed in the Hall fluid and correspondingly $\Psi_{\Lambda_{j}}(P^{\mathbf{i}})$ refers to the wave function of the $j^{th}$ particle located at the position $P^{\mathbf{i}}$. Now let us calculate the two-point correlation function in this fluid in the presence of only a $U(1)$ background. For a one-particle wave function in \eqref{oneparticlewavefunction} (with $n=1$) our notation transcribes as
\be
\Psi_{\Lambda_{i}}(P^{\mathbf{i}}) \equiv \Psi_{\alpha \beta}^{\mathbf{i}} \sim P_{\alpha \beta}^{\mathbf{i}}\,.
\label{U(1)func}
\ee 
The LLL wave function given in \eqref{eq:LLLwf} may now be denoted by
\be
\Psi_{\Lambda_{i}}(P^{\mathbf{i}}) \equiv \Psi_{\Lambda_{i}}^{\mathbf{i}} \sim (P_{\alpha \beta}^{\mathbf{i}})^n \,.
\label{U(1)funcn}
\ee
The general form of the correlation function between a pair of particles, say $1$ and $2$, on a manifold $\mathcal{M}$ is given by
\beq
\Omega(1,2) = \int_{\mathcal{M}}|\Psi_{MP}|^2 d\mu(3)d\mu(4)\cdots d\mu(\mathcal{N})\,,
\label{correlationfunc}
\eeq
with $d\mu(i)$ being the measure of integration on $\mathcal{M}$ in the coordinates of the $i^{th}$ particle and $\Psi_{MP}$ represents the multi-particle wave function of the Hall fluid on the manifold ${\cal M}$. Expanding the determinant formula \eqref{DetPsi} and using some algebra one can show that $\Omega(1,2)$ can be simplified as
\beq
\Omega(1,2)=\int_{\mathcal{M}}|\Psi_{MP}|^2 d\mu(3)d\mu(4)\cdots d\mu(\mathcal{N})=|\Psi^{\mathbf{1}}|^2|\Psi^{\mathbf{2}}|^2-|\Psi_\Lambda^{*\mathbf{1}}\Psi_\Lambda^{\mathbf{2}}|^2.
\label{correlationfuncmod}
\eeq
In order to compute \eqref{correlationfuncmod} for our case, we take the normalized coordinate chart $\gamma_i : = \frac{P_{\alpha \beta}}{P_{12}}$ where $P_{12} \ne 0$
\be 
\mathcal{P} = \frac{1}{\sqrt{1+|\gamma_{a}|^{2}}}(1,\gamma_{1},\dots,\gamma_{5})^T := \frac{1}{\sqrt{1+|\gamma_{a}|^{2}}} (1 \,, \vec{\gamma}) \,,
\ee
on the Grassmannian $\mathbf{Gr}_2(\mathbb{C}^4)$. In this coordinate patch \eqref{U(1)funcn} becomes $\Psi_{\alpha}^{\mathbf{i}} \sim (\mathcal{P}_{\alpha}^{\mathbf{i}})^n$. Inserting this into \eqref{correlationfuncmod} yields
\beqa
\Omega(1,2)&=& 1-|\mathcal{P}^{*\mathbf{1}}_\Lambda \mathcal{P}^{\mathbf{2}}_\Lambda|^{n} \nn \\
&=& 1-\left[\frac{\gamma_{a}^{*\mathbf{1}}\gamma_{a}^{\mathbf{2}}\gamma_{b}^{\mathbf{1}}\gamma_{b}^{*\mathbf{2}}}{1+\left|\gamma_{a}^{\mathbf{1}}\right|^{2}+\left|\gamma_{a}^{\mathbf{2}}\right|^{2}+\left|\gamma_{a}^{\mathbf{1}}\right|^{2}\left|\gamma_{a}^{\mathbf{2}}\right|^{2}}\right]^{n}\nn \\
&=& 1-\left[1-\frac{\left|\vec{\gamma}^{\mathbf{1}}-\vec{\gamma}^{\mathbf{2}}\right|^{2}}{1+\left|\gamma_{a}^{\mathbf{1}}\right|^{2}+\left|\gamma_{a}^{\mathbf{2}}\right|^{2}+\left|\gamma_{a}^{\mathbf{1}}\right|^{2}\left|\gamma_{a}^{\mathbf{2}}\right|^{2}}\right]^{n}.
\label{correlationfuncmod1}
\eeqa
Let us set $\vec{X}=\vec{\gamma}\ell$. In the thermodynamic limit ${\cal N} \rightarrow \infty$ and $n \rightarrow \infty$,
\eqref{correlationfuncmod1} takes the form
\beqa
\Omega(1,2) &=& 1-\left[1-\left|\vec{X}^{\mathbf{1}}-\vec{X}^{\mathbf{2}}\right|^{2}\left[\ell^{2}+\left|\vec{X}^{\mathbf{1}}\right|^{2}+\left|\vec{X}^{\mathbf{2}}\right|^{2}+\ell^{-2}\left|\vec{X}^{1}\right|^{2}\left|\vec{X}^{\mathbf{2}}\right|^{2}\right]^{-1}\right]^{n}\nonumber\\
&\rightarrow&1-\left[1-\tfrac{2B}{n}\left|\vec{X}^{\mathbf{1}}-\vec{X}^{\mathbf{2}}\right|^{2}\right]^{n}\nonumber\\
&{\rightarrow}&1-e^{-2B\left|\vec{X}^{\mathbf{1}}-\vec{X}^{\mathbf{2}}\right|^{2}} \nonumber\\
&=& 1-e^{-2B\left(\vec{x}^{\mathbf{1}}-\vec{x}^{\mathbf{2}}\right)^{2}}e^{-2B \ell^2\left({\det\Gamma}^{\mathbf{1}}-{\det\Gamma}^{\mathbf{2}}\right)^{2}},
\label{corr}
\eeqa
where we have used $n = 2 B \ell^2$ and that $\tiny{\Gamma^{\bf i} := \left(\begin{array}{cc}
\gamma_{2}^{\mathbf{i}} & \gamma_{1}^{\mathbf{i}}\\
\gamma_{4}^{\mathbf{i}} & \gamma_{3}^{\mathbf{i}}
\end{array}\right)}$. Note that the last equality shows the two-point function of the particles located at the positions $\vec{x}^{\bf 1},\vec{x}^{\bf 2}$ on $\mathbf{Gr}_2(\mathbb{C}^4)$, is extracted from that of the particles on $\mathbb{C}P^5$ at the positions $\vec{X}^{\bf 1},\vec{X}^{\bf 2}$ by a restriction of these particles to the algebraic variety determined by $X_5^{\bf i} \equiv \ell \det\Gamma^{\bf i}$, as expected. It is apparent from this function that the probability of finding two particles at the same point goes to zero. This result indicates the incompressibility of the Hall fluid.

Turning our attention to the $U(1)$ gauge field we may write  
\begin{equation}
A= - \frac{i n}{\sqrt{2}} \mathrm{Tr}\left(\lambda^{15}_{(6)}g^{-1} d g\right) \,.
\label{eq:9}
\end{equation}
With the help of (\ref{eq:su4par}) and (\ref{eq:r15}), one can express $A$ in terms of the Pl\"ucker coordinates as
\begin{align}
A & = -  \frac{i n}{\sqrt{2}}\left(\lambda_{(6)}^{15}\right)_{LM}\left(g^{-1}\right)_{MN}\left(dg\right)_{NL} \nonumber\\
 & =  -  \frac{i n}{2}\left(- \left(g^{-1}\right)_{5N}\left(dg\right)_{N5} + \left(g^{-1}\right)_{6N}\left(dg\right)_{N6}\right ) \nonumber\\
 & =   -\frac{i n}{2}\left(- g_{N5}^{*}\left(dg\right)_{N5} + g_{N6}^{*}\left(dg\right)_{N6}\right) \nonumber\\
 & =  -  \frac{i n}{2} \left(- P_{N} d P_{N}^{*} + P_{N}^{*} d P_{N} \right) \nonumber\\
 & =  - i n P_{N}^{*} d P_{N} \,,
\end{align}
where use has been made of the notational conventions stated below equation \eqref{eq:su4par}, and the fact that $d (P^*_N P_N) = 0$ due to \eqref{eq:pluckerrel}. Under $U(1)$ gauge transformations $A$ transforms to $A + d \left ( \frac{n \theta}{\sqrt{2}} \right )$, which is consistent with the transformation of the wave functions given in \eqref{eq:sections}. 

Let us introduce the notation $\mathcal{{\tilde P}} \equiv (P_{1},\dots P_6)^T$ where $T$ stands for transpose and define a non-homogeneous coordinate chart $\mathcal{Q} \equiv \frac{\mathcal{{\tilde P}}}{P_{1}}$ with $P_{1}\ne0$ on $\mathbf{Gr}_2(\mathbb{C}^4)$ as 
\begin{equation}
\mathcal{Q} \equiv (1\,,\gamma_{1}\,,\cdots\,,\gamma_5)^T \,,
\label{eq:pluckerq}
\end{equation}
subject to the Pl\"ucker relation (\ref{eq:pluckerrel}) which in terms of the (affine coordinates) $\gamma_i$ takes the form 
\be
\gamma_{5}=\gamma_{2}\gamma_{3}-\gamma_{1}\gamma_{4} \,.
\label{eq:pluckergamma}
\ee 
Without (\ref{eq:pluckergamma}), $\mathcal{Q}$ is a non-homogenous coordinate chart in ${\mathbb C}P^5$.
We can express our gauge potential as
\begin{eqnarray}
A &=&-in\mathcal{P}^{\dagger}d\mathcal{P} \nn \\
&=& -in|P_{1}|^{2}\mathcal{Q}^{\dagger}d\mathcal{Q}-inP_{1}^{*}|\mathcal{Q}|^{2}dP_{1}\nonumber\\
&=& -in|\mathcal{Q}|^{-2}\mathcal{Q}^{\dagger}d\mathcal{Q}-inP_{1}^{*}|P_{1}|^{-2}dP_{1}\nonumber\\
&=& -in|\mathcal{Q}|^{-2}\mathcal{Q}^{\dagger}d\mathcal{Q}-inP_{1}^{-1}dP_{1}\nonumber\\
&=& -in\partial\ln(|\mathcal{Q}|^{2})-ind\ln(P_{1})\nonumber\\
& = & -in\partial K-ind\ln(P_{1}).
\label{eq:AQ}
\end{eqnarray}
where $K$ is the ${\mathbb C}P^5$ K\"ahler potential given by 
\begin{equation}
K =\ln| \mathcal{Q}|^{2}\equiv\ln(1+|\gamma_{i} |^{2} ) \,,
\label{FSkahlerpotential}
\end{equation}
and subject to the condition (\ref{eq:pluckergamma}).

The field strength is calculated via
\beqa
F = d A &=&  - \frac{i n}{\sqrt{2}} \mathrm{Tr} \left ( \lambda^{15}_{(6)} g^{-1} d g \wedge  g^{-1} d g \right ) \nn \\
&=&  - i n  dP_{N}^{*} \wedge d P_{N} \,.
\eeqa
We note that $F$ is an antisymmetic, gauge invariant, and closed two-form on $\mathbf{Gr}_2(\mathbb{C}^4)$ and as such it is proportional to the K\"ahler two-form $\Omega$ over $\mathbf{Gr}_2(\mathbb{C}^4)$. This fact can be readily verified using (\ref{eq:AQ}) and writing
\begin{equation}
F = dA =  i n \partial {\partial}^* K = n \Omega \,,\label{F}
\end{equation}
where $\partial,{\partial}^*$ are the Dolbeault operators in the coordinates $\gamma_i$ and ${\gamma}^{*}_i$, respectively, and $d = \partial + {\partial}^*$. 
The relation \eqref{F} with \eqref{FSkahlerpotential} leads to the following form of the field strength \cite{Harris}:
\be
F = - i n \left ( \frac{d \gamma_i^* \wedge d \gamma_i }{1+|\gamma|^2} - \frac{ \gamma_i d \gamma_i^* \wedge \gamma_j^* d \gamma_j } {(1+|\gamma|^2)^2} \right ) \,,
\ee
being subject to the Pl\"ucker relation (\ref{eq:pluckergamma}). Let us associate with each index $i$ a dual index $\hat{i}$ in the sense that $i$ is dual to $\hat{i}$ if $\gamma_{i}\gamma_{\hat{i}}$ appears in the Pl\"ucker relation. Hence $1,4$ and $2,3$ are dual to one another. Expanding $\gamma_{5}$ in (\ref{eq:pluckergamma}) results in the  Hermitian components for the K\"ahler form
$\Omega$ as
\beqa
\Omega_{ii^{*}} & = & i N_{\gamma}\left(1+\overset{4}{\underset{\alpha=1,\alpha\ne i,\hat{i}}{\prod}}|\gamma_{\alpha}|^{2}+(1+|\gamma_{\hat{i}}|^{2})\overset{4}{\underset{\alpha=1,\alpha\ne i}{\sum}}|\gamma_{\alpha}|^{2}\right) \,, \nn \\
\Omega_{ij^{*}} & = & - i N_{\gamma}\left ( 1 + |\gamma_{\hat{i}}|^{2} + |\gamma_{\hat{j}}|^{2} \right ) \left ( \gamma_i^{*} \gamma_j + \gamma_{\hat{i}} \gamma_{\hat{j}}^{*} \right ) \,, \quad i < j \,, \quad j \neq {\hat i} \,, \\
\Omega_{i {\hat i}^{*}} & = & - i N_{\gamma} \left( \gamma_i^{*} \gamma_{\hat{i}} \left ( \overset{4}{\underset{\alpha=1}{\sum}}|\gamma_{\alpha}|^{2} - |\gamma_i |^{2} - |\gamma_{\hat{i}}|^{2} \right ) - \frac{1}{2} (\gamma_i^{*})^2 \prod_{j \neq i \,, {\hat i}} \gamma_j \gamma_{\hat j} - \frac{1}{2} (\gamma_{\hat i})^2 \prod_{j \neq i \,, {\hat i}} \gamma_j^{*} \gamma_{\hat j}^{*} \right ) \,, \quad i < {\hat i} \,, \nn
\eeqa
where $N_{\gamma}=\left(1+\sum_{a=1}^{5}|\gamma_{a}|^{2}\right)^{-2}$. In these formulas Einstein summation convention is not in use.

It is known from very general considerations \cite{Nair-book} that  the integral of $F$ over a non-contractable two surface $\Sigma$ in $\mathbf{Gr}_2(\mathbb{C}^4)$ is an integral multiple of $2 \pi$:  
\be
\frac{1}{2\pi}\int_\Sigma F =  n \,.\label{1ch}
\ee
In the present context,  this result signals an analogue of the Dirac quantization condition with $\frac{n}{2}$ identified as the magnetic monopole charge. Therefore, we do have that the magnetic field is $B = \frac{n}{2 \ell^2}$.

A number of remarks are in order. The generalization of our results to all higher dimensional Grassmannians is fairly straightforward. Taking $\mathbf{Gr}_2(\mathbb{C}^N)$, the only difference is that now both the vector potential $A$ and field strength $F$ are subject to the Pl\"ucker relations 
\be
\gamma_{ik}\gamma_{jl}=\gamma_{ij}\gamma_{kl}-\gamma_{il}\gamma_{kj},\quad 1\leq i<k<j<l\leq 2(N-2)\,,
\ee
in terms of the non-homogeneous coordinates $\gamma_{ij}:=P_{ij}/P_{12}$ in the patch where $P_{12}\ne0$. The parametrization in \eqref{eq:su4par} can be generalized to $N(N-1)/2$-dimensional fundamental representations of the $SU(N)$ group by means of these Pl\"ucker relations. Let us also note that the Grassmannians have a non-trivial algebraic topological structure that, for the best of our purposes here, is reflected in their second cohomology group which is non-zero, or more precisely $H^2(\mathbf{Gr}_k(\mathbb{C}^N)) = \mathbb{Z}$ \cite{Karshon}. This is the reason why the integral of the first Chern character in \eqref{1ch} is an integer. Similarly, one may consider the integral of the $d^{th}$ $(d=2(N-2))$ order Chern character for the Grassmannians $\mathbf{Gr}_2(\mathbb{C}^N)$ \cite{Gray}:
\be
\frac{1}{d!(2\pi)^{d}{\text {vol}(\mathbf{Gr}_2(\mathbb{C}^N))}}\int_{\mathbf{Gr}_2(\mathbb{C}^N)} F\wedge\Omega\cdots\wedge\Omega = n \,, \ee
for $F = n \Omega$.

\section{Conclusions and Outlook}

In this paper we have given a formulation of the QHE on ${\mathbf{Gr}_2(\mathbb{C}^N)}$. We solved the Landau problem on $\mathbf{Gr}_2(\mathbb{C}^N)$ using group theoretical techniques and gave the energy spectra and the wave functions of charged particles on $\mathbf{Gr}_2(\mathbb{C}^N)$ in the background of both abelian and non-abelian magnetic monopoles. For abelian monopole background, using the local description of wave functions in terms of Pl\"{u}cker coordinates on ${\mathbf{Gr}_2(\mathbb{C}^4)}$, we showed that the LLL at filling factor $\nu =1$ forms an incompressible fluid and indicated how this result generalizes to all ${\mathbf{Gr}_2(\mathbb{C}^N)}$.

We want to make the following observations about the QHE on $\mathbf{Gr}_2(\mathbb{C}^4)$ with $U(1)$ background. Due to the isomorphisms $Spin(6) \cong SU(4)$ and $Spin(4) \cong SU(2) \times SU(2)$, the Stiefel manifold $\mathbf{St}_2(\mathbb{R}^6) \equiv \frac{Spin(6)}{Spin(4)}$ forms the principal $U(1)$ fibration \cite{Uskov}
\be
U(1) \longrightarrow  \mathbf{St}_2(\mathbb{R}^6) \longrightarrow \mathbf{Gr}_2(\mathbb{C}^4) \,.
\ee
Let us also make note of the family of fibrations $ \mathbf{St}_{k-1}(\mathbb{R}^{n-1}) \longrightarrow \mathbf{St}_{k}(\mathbb{R}^n) \longrightarrow S^{n-1}$, which for $k=2$ and $n=6$ is
\be
S^4 \longrightarrow \mathbf{St}_{2}(\mathbb{R}^6) \longrightarrow S^{5} \,.
\ee
Together these facts imply that $\mathbf{Gr}_2(\mathbb{C}^4)$ has the local structure $\frac{S^5 \times S^4}{U(1)}$. We therefore propose  that the QHE on $S^5$ with the $S^4$ fibers may be seen as a QHE on $\mathbf{Gr}_2(\mathbb{C}^4)$ with a $U(1)$ background gauge field.
We expect that the $S^4$ fibers will be associated to a $SO(5)$ gauge field background. In fact, the formulation of the QHE on the $3$-sphere \cite{Nair-Daemi}
\be
S^3 = \frac{SU(2) \times SU(2)}{SU(2)_{diag}} \cong \frac{Spin(4)}{Spin(3)} \,,
\ee
selects the constant background gauge field as the spin connection and in a construction generalizing this to the QHE on $S^5$,
\be
S^5 = \frac{SO(6)}{SO(5)} = \frac{Spin(6)}{Spin(5)} \,,
\ee
one should be selecting a constant $SO(5)$ background gauge field taking it again as the spin connection. Such a choice of the gauge field appears to be consistent with our heuristic argument. Our observation is inspired by and bears a resemblance to the relation between the QHE on ${\mathbb C}P^7$ and $S^8$. The former can be realized locally as $\frac{S^8 \times S^7}{U(1)}$, while the latter forms the base of the $3^{rd}$ Hopf map $S^7 \longrightarrow S^{15} \longrightarrow S^8$, and $S^{15}$ is a $U(1)$ bundle over ${\mathbb C}P^7$ \cite{Bernevig-S8}.

A number of future directions for further research is foreseen. Firstly, the aforementioned relation with the QHE on $S^5$ could be made more concrete by developing the latter along the lines discussed here and adapting the approach of \cite{Bernevig-S8} and \cite{Nair-Daemi}. It is known that the LLL wave functions for the QHE on ${\mathbb C}P^N$ has a correspondence with the algebra of functions on fuzzy ${\mathbb C}P^N$ \cite{Nair-Daemi-Karabali}. Extending these results to the LLL wave functions on $\mathbf{Gr}_2(\mathbb{C}^N)$ and fuzzy Grassmannians as discussed in \cite{Dolan, Chris, Huet-Murray}, would provide additional insights. It may also be possible to develop Chern-Simons type effective field theories along the lines of \cite{Bernevig-EFT} to shed more light on the structure of the QHE on $\mathbf{Gr}_2(\mathbb{C}^4)$ in particular. Formulation of the edge states may also be investigated building upon the ideas of \cite{Nair2, Nair-Karabali-Effective2}. We hope to report on the progress of any of these topics elsewhere.

\vskip 2em

\noindent {\bf \large Acknowledgements}

\vskip 1em

We would like to thank  A.P. Balachandran and C.\"{U}nal for useful discussions. A. B. is thankful to E. Co\c{s}kun and M. Bhupal for valuable discussions. F. B., S. K., and G.\"{U}. acknowledge the support of T\"{U}B\.{I}TAK under the project No.110T738. S.K. also acknowledges the support of T\"{U}BA-GEBIP program of The Turkish Academy of Sciences. 

\vskip 1em

\appendices

\subsection{Appendix A}

In this short appendix, we provide a derivation of the normalization coefficient of $R_{N^2-1}$ in the $\frac{N(N-1)}{2}$-dimensional IRR of $SU(N)$ for $N \geq 3$. Let $T^{(D)}_a$ label the $N^2-1$ generators of $SU(N)$ in the defining $N$-dimensional representation. Let us choose their trace normalization to be
\be
\mbox{Tr}(T^{(D)}_a T^{(D)}_b) = \frac{1}{2} \delta_{ab} \,.
\ee
It is a well-known fact in the representation theory of Lie groups that such a choice fixes the trace normalization of the generators in all the IRR \cite{Fuchs}. We can proceed to write the trace normalization in an IRR $R$ of $SU(N)$ as 
\beq
\mbox{Tr} (T^{(R)}_aT^{(R)}_b) = \kappa_{ab} \,, 
\label{trace_R}
\eeq 
where $\kappa_{ab}$ is a rank-$2$ tensor invariant under $SU(N)$ transformations. Since the only rank-$2$ invariant $SU(N)$ tensor is Kronecker delta, $\delta_{ab}$, we have 
\beq
\kappa_{ab}=X_{(R)} \delta_{ab} \,,
\eeq 
where $X_{(R)}$, commonly known as the {\it Dynkin index} of the representation $R$ of the group $SU(N)$, is given by
\cite{Fuchs}
\beq
X_{(R)}=\frac{\dim(R)}{\dim(SU(N))}\mathcal{C}_2(R) \,.
\eeq 
We have that $\dim(SU(N))$ is equal to $N^{2}-1$ and $\mathcal{C}_2^R$ is the quadratic Casimir of the IRR $R$. For either of the $\frac{N(N-1)}{2}$-dimensional IRR, $(0\,,1\,,0\,,\cdots\,, 0\,, 0)$ or $(0\,,0\,,\cdots\,,1\,, 0)$ of $SU(N)$, this gives, using (\ref{eq:casuN}),
\begin{equation}
X_{(R)}=\frac{N-2}{N} \,,
\end{equation}
and the trace formula \eqref{trace_R} then reads
\beq
\mbox{Tr} (T_aT_b) =\frac{N-2}{N} \delta_{ab}, 
\label{eq:traceR2} 
\eeq 
in either of the $\frac{N(N-1)}{2}$-dimensional IRR. Our aim is to find the coefficient of $R_{N^2-1}$ in these representations. In terms of the Young diagrams, the branching of, say, $(0\,,1\,,0\,,\cdots\,, 0\,, 0)$ representation under $SU(N-2) \times SU(2) \times U(1)$ gives 
\begin{equation}
\label{eq:branching(0,1,...,0)}
\smng(1,1)~\ =\ ~
\bigg(\cdot~\otimes~\smng(1,1)~\bigg)_{-1}\oplus~\bigg(~\smng(1,1)~\otimes~\cdot\bigg)_{\frac{2}{N-2}}
~\oplus~\bigg(~\smng(1)~\otimes~\smng(1)~\bigg)_{\frac{4-N}{2(N-2)}},
\end{equation}
where the subscripts give the $U(1)$ charge (\ref{eq:charge}). Considering the dimension of each representation in this branching, we find 
\beq
R_{N^2-1}= \zeta \, \text{diag}\bigg (\underbrace{\frac{N-4}{2(N-2)},\dots,\frac{N-4}{2(N-2)}}_{2(N-2)} \,, \underbrace{\frac{-2}{N-2},\dots,\frac{-2}{N-2}}_{\frac{(N-2)(N-3)}{2}} \,, 1 \bigg ) \,, 
\label{u(1)charge_dim}
\eeq
where $\zeta$ represents the coefficient of $R_{N^2-1}$ and the dimensions of the IRR in the branching \eqref{eq:branching(0,1,...,0)} are given in the underbraces. Finally, using \eqref{u(1)charge_dim} in (\ref{eq:traceR2}) gives 
\begin{equation}
\zeta=\sqrt{\frac{N-2}{N}} \,.
\end{equation}

\setcounter{equation}{0}

\subsection{Appendix B}

The dimension of the $(P_1\,, P_2\,,P_3\,,\cdots\,, P_{N-2} \,, P_{N-1})$ representation may be written as
\begin{eqnarray}\label{eq:Dimension_of_generic_SU(N)}
\dim(P_1,P_2,0,\dots,0,P_{N-2},P_{N-1}) & = &  \frac{1}{j}((P_{N-2}+P_{N-1}+N-3)! (P_{N-2}+N-4)! (P_{2}+N-4)!\nonumber\\&\times&(P_{1}+P_{2}+N-3)!(P_{N-2}+P_{N-1}+P_{2}+N-2)\nonumber\\&\times&(P_{N-1}+1)(P_{1}+P_{2}+P_{N-2}+P_{N-1}+N-1)(P_{1}+1)\nonumber\\&\times&(P_{N-2}+P_{2}+N-3)(P_{1}+P_{2}+P_{N-2}+N-2)),
\label{eq:dimsuN}
\end{eqnarray}

where $j$ is 
\begin{equation}
j=(N-1)!(N-2)!(N-3)!(N-4)!P_{2}!P_{N-2}!(P_{N-2}+P_{N-1}+1)!(P_1+P_2+1)!.
\end{equation}

\end{document}